\documentclass{aa}
\usepackage{graphics}

\def\cc{\,{\rm cm^{-3}}}
\def\cm2{\,{\rm cm^{-2}}}
\def\pc2{\,{\rm pc^{2}}}
\def\kms{\,{\rm {km\,s^{-1}}}}
\def\kkms{\,{\rm {K\,km s^{-1}}}}
\def\co{\,{\rm ^{12}CO}}
\def\13co{\,{\rm ^{13}CO}}
\def\h2{\,{\rm H_{2}}}

\def\cii{[C$\;$II]}
\def\hii{H$\;$II}

\def\nhmin{{N_H^{\rm min}}}
\def\mhmin{{M_H^{\rm min}}}
\def\go{{G_o}}
\def\fuv{{f_{\rm UV}}}

\def\fcii{{F_{\rm C\;II}}}
\def\fco{{F_{\rm CO}}}
\def\fir{{F_{\rm IR}}}

\def\Icii{{I_{\rm C\;II}}}
\def\Iir{{I_{\rm IR}}}
\def\taucii{{\tau_{\rm C\;II}}}
\def\texc{{T_{\rm ex}}}
\def\xcp{{x_{\rm C^+}}}
\def\xc{{x_{\rm C}}}
\def\cp{{\rm C^{+}}}
\def\aeff{{A_{\rm eff}}}
\def\mvt{{M_{\rm VT}}}
\def\mpdr{{M_{\rm PDR}}}
\def\mm{{\rm $\mu$m}}
\def\co10{CO($1\rightarrow 0$)}
\def\coa32{CO($3\rightarrow 2$)}
\def\13co32{$^{13}$CO($3\rightarrow 2$)}
\def\twelveco{$^{12}$CO}
\def\cob43{CO($4\rightarrow 3$)}
\def\coc54{CO($5\rightarrow 4$)}
\def\cod76{CO($7\rightarrow 6$)}

\def\hcna10{HCN($1\rightarrow 0$)}
\def\hcnb43{HCN($4\rightarrow 3$)}
\def\car10{CI($^3P_1\rightarrow ^3P_0$)} 
\def\ioncar{$^2P_{3/2}\rightarrow ^2P_{1/2}$} 
\def\carb21{CI($^3P_2\rightarrow ^3P_1)$} 
\def\etal{et al.\ }
\def\eg{e.g.,\ }
\def\ie{i.e.,\ }
\def\psqcm{\ifmmode {\>{\rm cm}^{-2}}\else {cm$^{-2}$}\fi}
\def\pcubcm{\ifmmode {\>{\rm cm}^{-3}}\else {cm$^{-3}$}\fi}
\def\be{\begin{equation}}
\def\ee{\end{equation}}
\def\bea{\begin{eqnarray}}
\def\eea{\end{eqnarray}}
\def\rpcsq{\ifmmode {r_{\rm pc}^2}\else {$r_{\rm pc}^2$}\fi}
\def\rpc{\ifmmode {r_{\rm pc}}\else {$r_{\rm pc}$}\fi}
\def\fabs{\ifmmode {f_{\rm abs}}\else {$f_{\rm abs}$}\fi}
\def\msol{\ifmmode {\>M_\odot}\else {$M_\odot$}\fi}
\def\lsol{\ifmmode {\>L_\odot}\else {$L_\odot$}\fi}
\def\ergsr{\ifmmode {\rm\> erg\;cm^{-2}\;s^{-1}\;sr^{-1}} 
            \else {erg cm$^{-2}$ s$^{-1}$ sr$^{-1}$}\fi}
\def\ergs{\ifmmode {\rm\> erg\;cm^{-2}\;s^{-1}} 
            \else {erg cm$^{-2}$ s$^{-1}$}\fi}
\def\aua{{\rm A\&A} }
\def\auas{{\rm A\&AS} }

\def\apj{{\rm ApJ} }
\def\aj{{\rm AJ} }
\def\apjs{{\rm ApJS} }
\def\apjl{{\rm ApJL} }
\def\araa{{\rm ARAA} }
\def\mnras{{\rm MNRAS} }

\begin{document}

\title{C$^{+}$ Emission from the Magellanic Clouds} 

\subtitle{II. [CII] maps of star-forming regions LMC-N~11, SMC-N~66, and
  several others}

\author{F.P. Israel \inst{1}
    and P.R. Maloney \inst{2}	}

\offprints{F.P. Israel}
 
  \institute{Sterrewacht Leiden, Leiden University, P.O. Box 9513,
    2300 RA Leiden, The Netherlands 
       \and Center for Astrophysics and Space Astronomy, University of
    Colorado, Boulder, CO 80303, USA  }

\authorrunning{F.P. Israel $\&$ P.R. Maloney}

\titlerunning{C$^{+}$ in the Magellanic Clouds}

\date{Received ????; accepted ????}
 
\abstract{We study the $\lambda$ 158 $\mu$m [CII] fine-structure line
  emission from star-forming regions as a function of metallicity.}
         {We have measured and mapped the [CII] emission from the very
           bright HII region complexes N~11 in the LMC and N~66 in the
           SMC. as well as the SMC \hii\ regions N~25, N~27,
           N~83/N~84, and N~88 with the FIFI instrument on the Kuiper
           Airborne Observatory.}  {In both LMC and SMC, the ratio of
           \cii\ line to CO line and to the far-infrared continuum
           emission is much higher than seen almost anywhere else,
           including Milky Way star-forming regions, and whole
           galaxies.}  {In the low metallicity, low dust-abundance
           environment of the LMC and the SMC UV mean free path
           lengths are much greater than those in the
           higher-metallicity Milky Way. The increased photoelectric
           heating efficiencies cause significantly greater relative
           \cii\ line emission strengths. At the same time, similar
           decreases in PAH abundances have the opposite effect by
           diminishing photoelectric heating rates.  Consequently, in
           low-metallicity environments relative \cii\ strengths are
           high but exhibit little further dependence on actual
           metallicity.  Relative \cii\ strengths are slightly higher
           in the LMC than in the SMC which has both lower dust and
           lower PAH abundances.}  {}{}

\maketitle

\section{Introduction}
\begin{table*}
\scriptsize
\caption[]{\cii\ Observations}
\begin{center}
\begin{tabular}{lccccccc}
\noalign{\smallskip}     
\hline
\noalign{\smallskip}
Field &\multicolumn{2}{c}{Reference Position}& Integration& 1$\sigma$ Noise& Number& Field\\
& RA(1950)   & DEC (1950)                    & Time       & Level          & of Array&Name\\
& (h m s) & ($^{o}$ $'$ $''$)                 & (min)      & ($\ergsr$)     & Pointings& \\
\noalign{\smallskip}     
\hline
\noalign{\smallskip}     
1  & 00 46 20.0 & -73 31 10  &  7  & 0.90 (-5) &  1  & N25 \\
2  & 00 46 30.0 & -73 22 00  &  8  & 0.83 (-5) &  4  & N27 \\
3  & 00 57 40.0 & -72 25 00  & 10  & 0.68 (-5) &  5  & N66 \\
4  & 01 12 30.0 & -73 33 00  &  5  & 0.94 (-5) &  7  & N83/N84 \\
5  & 01 22 54.0 & -73 24 30  & 12  & 0.63 (-5) &  1  & N88 \\
6  & 04 57 00.0 & -66 30 00  &  5  & 0.94 (-5) & 18  & N11 \\ 
\noalign{\smallskip}     
\hline
\noalign{\smallskip}
\end{tabular}
\end{center}
\label{kaofields}
%Notes to Table 1:\\
\end{table*}

Emission lines from carbon and carbon monoxide provide almost all the
cooling of dense neutral gas in the interstellar medium.  With a
critical density for excitation $\sim 3\times10^3 \cc$, the
\ioncar\ fine-structure line of singly ionised carbon ($\cp$) at 157.7
\mm\ is a major cooling line for regions exposed to significant FUV
photon fluxes (photon-dominated regions or PDRs: Tielens \& Hollenbach
1985a,b). In Galactic \hii\ regions as well as in the central regions
of external galaxies (for instance Howe \etal 1991; Stacey \etal 1991;
Jaffe \etal 1994; Malhotra \etal 2001; Negishi \etal 2001) the
luminosity of the \cii\ line is typically $\sim 0.05-0.5\%$ of the
far-infrared luminosity ($FIR$) and correlates well with CO line
intensities.  For the Milky Way as a whole, the COBE measurements
published by Wright \etal (1991) correspond to
$I_{{CII}}/FIR=0.1-0.2\%$.  Substantially higher \cii\ to far-infrared
ratios $I_{[CII]}/FIR=1.6-3.4\%$ occur in high-latitude translucent
clouds in the Milky Way (Ingalls, Reach \& Bania, 2002)

Most of the 158\mm\ observations published to date have sampled
objects with metal abundances similar to or greater than those in the
Solar Neighbourhood.  Moreover, they have usually sampled either
relatively small regions of space in our own Galaxy, or very large
volumes in other galaxies (typically the inner few kpc).  The nearby
Magellanic Clouds (LMC: 50 kpc, Schaefer 2008; SMC: 61 kpc, Szewczyk
\etal 2009) provide an ideal opportunity to study objects with
abundances significantly below solar, and at the same time study
intermediate spatial scales.  They are rich in interstellar gas and
young, luminous stars, but carbon abundances are particularly low. The
data reviewed by Pagel (2003) show that LMC and SMC HII regions have
abundances (12+lg[C]/[H] = 7.9 and 7.5, respectively) so that compared
to the Milky Way, the LMC is under-abundant in C by a factor of four
and the SMC by a factor ten.  Due to its reduced metallicity and dust
content, the neutral gas in the Clouds is substantially less shielded
from UV continuum radiation long-ward of the Lyman limit than the gas
in our own Galaxy. Because of its low ionisation potential of 11.3 eV,
neutral carbon is the most abundant heavy element that can be ionised
by the relatively unobstructed far-ultraviolet (FUV) radiation.

Mochizuki \etal (1994) have surveyed the whole LMC in the \cii\ line
at a relatively low resolution of 12.4' (corresponding to a linear
resolution of 194 pc) with the balloon-borne facility BICE.  Their
survey reveals a number of bright, discrete \cii\ sources in addition
to extended emission.  The two brightest sources coincide with the
very bright LMC \hii\ region 30 Doradus and the N~160/N~159 \hii\
complex to the south of it.  These have been mapped in \cii\ at much
higher resolution by Poglitsch \etal (1995) and Israel \etal (1996;
Paper I), respectively.  The study of these very bright LMC \hii\
region complexes has showed that they resemble Galactic translucent
clouds, with \cii\ line luminosities being a similarly large fraction
of the total far-infrared luminosity, typically $\sim 1 - 3\%$. The
\cii\ emission exhibits a poor correlation with CO line intensities on
the relevant 10 pc scales: the local $\fcii/\fco$ ratios vary from
$400$ (N159S) to $113,000$ (30 Dor).

We have shown in Paper I that at low to moderate densities ($n\approx
10^2 - 10^4\cc$), the \cii\ 158\mm\ line intensity varies roughly with
the incident far-UV radiation field for $\go$ = 1-100
\footnote{$\go$ is the radiation field intensity in units of $1.6
  \times 10^{-3}\,\ergs$ (\ie the strength of the Solar Neighbourhood
  interstellar radiation field: Habing 1968)}. The intensity of the
$J$=1-0 \twelveco\ line varies much more slowly with $\go$, but at
higher values of $\go$ the \cii\ line intensity saturates. In this
paper, we pursue these conclusions and present results obtained
towards N~11, the fourth brightest \cii\ source in the LMC (after
N~44), as well as a dozen \hii\ regions in the lower-metallicity SMC,
including the brightest SMC \hii\ region, N~66 (Henize 1956).

\section{Observations}
\nobreak The \cii\ 158\mm\ maps presented in this paper were made with
the MPE/UCB Far-Infrared Imaging Fabry-Perot Interferometer (FIFI;
Poglitsch \etal 1991) during several flights of the NASA Kuiper
Airborne Observatory (KAO) out of Christchurch (New Zealands) in April
1992.  FIFI had a $5\times 5$ focal plane array with detectors spaced
by $40''$ (Stacey \etal 1992). Each detector had a FWHM of $55''$
corresponding to 14.3 pc on the LMC and 16.2 pc on the SMC. The beam
shape was approximately gaussian ($68''$ equivalent disk; beam solid
angle $\Omega_B=8.3\times 10^{-8}$ sr). The Lorentzian instrumental
profile gave a 50 $\kms$ spectral resolution. To optimise our
sensitivity to extended low level emission, we observed in `stare'
mode, by setting the bandpass of the Fabry-Perot to the line-center at
the object velocity. To eliminate telescope offset and drift, all
observations were chopped at 23 Hz and beam-switched once a minute
against two reference positions about $6'$ away. IRAS maps were used
to verify that the reference positions were devoid of strong discrete
far-infrared sources and therefore could reasonably be expected to be
also free from discrete sources of contaminating line emission.  The
data were calibrated by observing an internal black-body source. The
calibration uncertainty is of order 30\% and the absolute pointing
uncertainty is well below $15''$.  The observed fields are summarised
in Table \ref{kaofields}, and the maps are shown in Figures 1 through
4.

In Table \ref{kaofields}, we list the position used as (0,0)
reference in the maps (columns 2 and 3), the integration time for each
array pointing (column 4), the resulting noise level in the maps
(column 5), the number of separate array pointings used in
constructing the maps (column 6), and the most prominent \hii\ region
in the field (column 7).  In all maps, the array pointings were
adjacent to one another. In SMC fields 3 and 4, we obtained additional
array pointings on the strongest \cii\ emission regions corresponding
to the location of the \hii\ regions SMC N~66, SMC N~83, and SMC N~84
in such a way as to obtain effective detector separations of $30''$ on
those regions.  The central part of the LMC N~11 complex was covered
by 18 different, partially overlapping array pointings.  The central
3$'$ $\times$ 3$'$ centered on LH~10 (N~11A and N~11B) was also
observed with two different array pointings, offset from one another
by half a beam separation on both the N-S and E-W directions.  This
field was therefore fully sampled, in contrast with the rest of the
map which is slightly under-sampled.

All maps contain one or more peaks of \cii\ emission. In Tables
\ref{lmcclouds} and \ref{smcclouds} we provide information on the
objects included in the fields mapped, and their association with
\cii\ peaks in the maps.  In Table \ref{lmcclouds}, columns 2 through
4 provide the Henize (1956) object name, the associated CO cloud from
Israel \etal (2003a), and the CO cloud offset position with respect to
the reference position in Table \ref{kaofields}.  Column 5 through 7
give the Lucke $\&$ Hodge (1970) OB association number, the IRAS
source name and position taken from Schwering $\&$ Israel (1990),
while columns 8 and 9 give the \hii\ region H$\alpha$ line flux
(Caplan \etal 1996) and the 4750 MHz radio continuum flux density
(Filipovi\'c \etal 1998).  In Table \ref{smcclouds}, columns 2 through
4 provide the Henize (1956), Davies \etal (1976), and NGC object
names.  Column 5 lists the \hii\ region position from Davies \etal
with respect to the reference position in Table \ref{kaofields}.
Columns 6 and 7 give the corresponding IRAS source name and offset
position taken from Schwering $\&$ Israel (1990), while columns 8 and
9 give the \hii\ region H$\alpha$ line flux (Caplan \etal 1996;
Cornett \etal 1997) and the 843 MHz radio continuum flux density (Ye
\etal 1991).  Detailed information for the \cii\ peaks themselves is
given in Tables \ref{lmcdat} and \ref{smcdat}.

\begin{table*}
\scriptsize
\caption[]{LMC-N~11 objects in [CII] map }
\begin{center}
\begin{tabular}{lcccccccc}
\hline
\noalign{\smallskip}     
CII Peak & \multicolumn{6}{c}{Sources Included} 	        & \multicolumn{2}{c}{Integrated Flux} \\
   & Henize & CO & CO Position & LH OB & IRAS   & IRAS Position & F(H$_{\rm \alpha}$) & S(4750) \\
   &        & Cloud & ($'$)    &     &          & ($'$)         & ($\ergs$)           & (mJy) \\
\noalign{\smallskip}     
\hline
\noalign{\smallskip}
1  & ---   &  5 &--9.0, +0.5 & (9)& LILMC195 &--8.2, --0.4 & ---       &  --- \\
2  & N11F  &  8 &--3.9, --6.4&  9 & LILMC214 &--2.5, --7.4 & 1.7 (-10) &  683 \\
3  & N11F  &  9 &--2.5, --7.0&  9 & idem     & 	     & included  & included   \\
4  & N11B  & -- & ---        & 10 & LILMC217 &--2.0, +1.0  & ---       &  --- \\
5  & N11AB & 10 &--1.5, +1.0 & 10 & LILMC226 & +0.9, +2.3  & 9.0 (-10) & 3722 \\
6  & ---   & 11 & +0.8, +2.0 & -- & ---      & ---         & ---       &  --- \\
7  & ---   & 12 & +0.8, +7.0 & -- & LILMC229 & +2.1, +6.1  & ---       &  --- \\
8  & ---   & 13 & +1.5, +3.0 & -- & LILMC226 & +0.9, +2.4  & ---       &  --- \\
9  & N11C  & 14 & +3.4, --1.3& 13 & LILMC243 & +3.6, --1.9 & 4.5 (-10) &  641 \\
10 & N11D  & 15 & +4.7, --3.3& 13 & LILMC248 & +4.0, --3.3 &included &included\\
11 & N11E &16-18& +5.1, +4.8 & 14 & LILMC251 & +6.5, +3.8  & 1.1 (-10) &  306 \\
\noalign{\smallskip}     
\hline
\end{tabular}
\end{center}
%Notes to Table 1:\\
\label{lmcclouds}
\end{table*}

\begin{table*}
\scriptsize
\caption[]{LMC-N~11 [CII] Data}
\begin{center}
\begin{tabular}{lcccccccc}
\hline
\noalign{\smallskip}     
[CII] & Peak Intensity & \cii\ Position   & \multicolumn{3}{c}{Integrated Flux} & \multicolumn{3}{c}{Ratios}   \\
Cloud & $I_{\rm CII}$ &               & $F_{\rm CII}$ &$F_{\rm CO}$ & $F_{\rm IR}$
& {$F_{\rm CII}\over F_{\rm CO}$} & {$F_{\rm CII}\over F_{\rm IR}$} &
{$F_{\rm IR}\over F_{\rm CO}$} \\
      & ($\ergsr$)     & ($'$)        & \multicolumn{3}{c}{($\ergs$)}       & & & \\
\noalign{\smallskip}     
\hline
\noalign{\smallskip}
1  & 7.8 (-5)  & -8.3, +0.7 & 3.1 (-11) & 0.6 (-15) & 0.9 (-9) & 5.2 (4) & 3.6 (-2) & 1.4 (6) \\
2  & 9.7 (-5)  & -3.7, -6.8 & 1.8 (-11) & 0.4 (-15) & 2.4 (-9) & 4.5 (4) & 1.8 (-2) & 6.0 (6) \\
3  & 9.4 (-5)  & -2.3, -6.8 & 2.4 (-11) & 0.4 (-15) & included & 6.0 (4) & included & included\\
4  & 20  (-5)  & -3.4, +1.7 & 4.6 (-11) & 1.3 (-15) & 15  (-9) & 3.5 (4) & 1.0 (-2) & 11  (6) \\
5  & 22  (-5)  & -1.7  +1.0 & 9.2 (-11) & included  & included & included& included & ---     \\
6  & 13  (-5)  & +0.8, +1.2 & 3.1 (-11) & 0.3 (-15) & ---      & 10  (4) &  ---     & ---     \\
7  & 2.6 (-5): & +1.7, +6.3 & 0.7 (-11) & 0.4 (-15) & ---      & 1.8 (4) &  ---     & ---     \\
8  & 6.8 (-5)  & +2.0, +4.5 & 1.9 (-11) & 0.3 (-15) & ---	& 6.3 (4) &  ---     & ---     \\
9  & 17  (-5)  & +3.0, -1.7 & 7.8 (-11) & 0.3 (-15) & 2.7 (-9) & 26 (4)  & 4.9 (-2) & 8.9 (6) \\
10 & 13  (-5)  & +3.7, -3.0 & 5.2 (-11) & 1.3 (-15) & included & 4.0 (4) & included & included\\
11 & 7.3 (-5)  & +6.0, +5.0 & 4.2 (-11) & 1.5 (-15) & 1.9 (-9) & 2.8 (4) & 2.3 (-2) & 1.2 (6) \\
\noalign{\smallskip}     
\hline
\end{tabular}
\end{center}
%Notes to Table 1:\\
\label{lmcdat}
\end{table*}

\section{Results and analysis}

\subsection{LMC}

\subsubsection{The N~11 Field}

%Figure 1  N11
\begin{figure*}[]
\unitlength1cm
\begin{minipage}[t]{18cm}
\resizebox{18.57cm}{!}{\rotatebox{270}{\includegraphics*{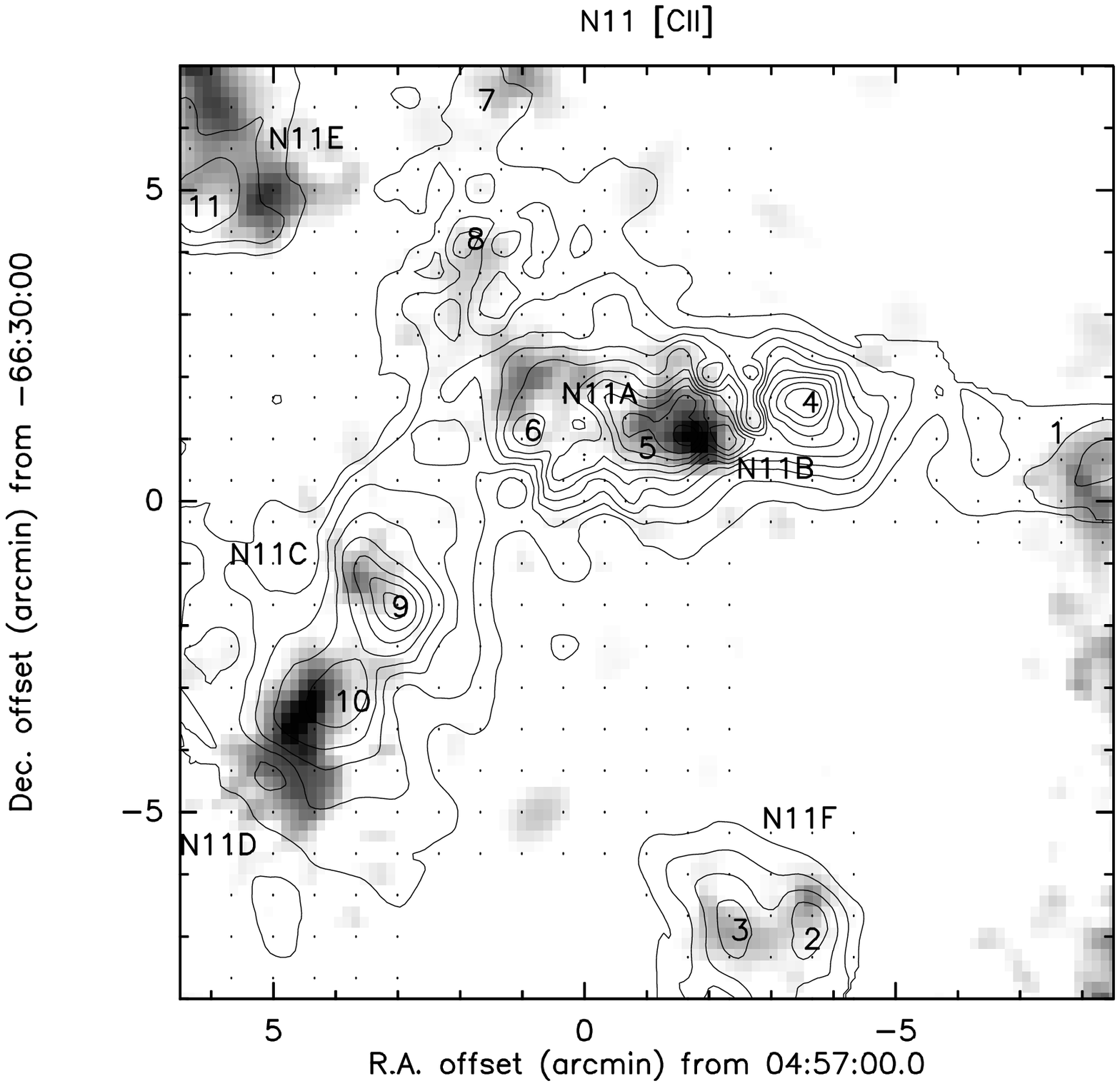}}}
\end{minipage}
\caption[]{Map of the central part of LMC-N~11. \cii\ contours are
  linear at multiples of $2.125\times10^{-5}\,\ergsr$.  Integrated
  $J$=1-0 $^{12}$CO emission is represented by grayscales in the range
  1-10 $\kkms$. Small dots mark the grid points sampled in [CII].  The
  locations of the major \hii\ region components N~11A through N~11F
  are schematically indicated. All \cii\ peaks (marked by their number
  from Tables 2 and 3) are associated with a CO cloud.  The lack of CO
  emission near \cii\ cloud 4 may reflect poor sampling of the CO map
  at this position. }
\label{N11map}
\end{figure*}

\nobreak The third most luminous star-forming complex in the LMC,
N~11, is located opposite 30 Doradus at the northwestern edge of the
galaxy.  Unlike the highly centralised 30 Doradus region, N~11 is an
extended complex of individual OB associations catalogued by Lucke
$\&$ Hodge (1970).  These illuminate a variety of discrete \hii\ and
CO clouds (Israel \etal 1993; Mac Low \etal 1998).  A striking feature
of N~11 is the super-bubble (Meaburn \etal 1989) formed by the action
of stellar winds from the OB association LH~9 (NGC~1760; age 7 Myr,
Mokiem \etal 2007) at its center. The bubble is a prominent source of
soft X-rays (Mac Low \etal 1998, Naz\'e \etal 2004, Maddox \etal 2009)
implying significant interaction between the stars and the surrounding
gas. At the northern edge of the bubble is the much younger OB
association LH~10 (NGC~1763; 3 Myr, Mokiem et al. 2007) which is still
embedded in the bright nebula N~11B (cf. Heydari-Malayeri \& Testor
1983). Three O3 stars have been identified in the LH~10 association
(Parker \etal 1992). An even younger compact object, N~11A (IC~2116;
Mac Low \etal 1998; Naz\'e \etal 2001; Heydari-Malayeri \etal 2001),
occurs in the same area. The location and apparent age difference of
LH~10 and LH~9, as well as the gas kinematics, have led to the
suggestion that the formation of LH~10 and other stellar groups was
triggered by the expanding shells emanating from LH~9 (Walborn $\&$
Parker 1992; Rosado \etal 1996; Barba \etal 2003; Hatano \etal 2006;
Mokiem \etal 2007).

The OB association LH~13 (NGC~1769), to the east of the bubble,
appears to be younger than 5 Myr and excites the bright \hii\ region
N~11C/D, an ionised region seemingly divided in two by a dust band
crossing in front (Heidari-Malayeri \etal 1987).  The association
LH~14 (NGC~1773), northeast of the bubble, excites the \hii\ region
N~11E and should also be young as it contains an O4-5V star
(Heidari-Malayeri \etal 1987).

Much of N~11 was surveyed in the $J$=1-0 transition of $^{12}$CO as
part of an ESO-Swedish Key Programme; more limited mapping of the
central part was conducted in the $J$=2-1 transition (Israel et
al. 2003a).  The N~11 molecular clouds are quite warm, with kinetic
temperatures of 60 - 150 K.  This is higher than expected for clouds
heated by stellar UV photons only. The CO clouds are also quite
distinct, i.e. there is almost no diffuse CO emission between the
bright clouds. Both aspects suggest a strong influence of the ambient
radiation field (UV, X-rays) on the molecular gas including processing
of the diffuse gas.

Our \cii\ line observations, although not fully covering the large
N~11 complex, have a spatial resolution similar to that of the
\twelveco\ maps of Israel \etal (2003a). In Fig.~\ref{N11map}
contours trace the distribution of \cii\ emission, and shades of gray
represent the $J$=1-0 CO distribution (cf Fig. 1 of Israel \etal
2003a). Peak and integrated \cii\ intensities of the various clouds
detected are listed in Table \ref{lmcdat}, Columns 2 and 4
respectively.

The CO and the \cii\ distributions are roughly similar. Both trace the
circumference of the super-bubble and follow the distribution of the OB
associations and \hii\ regions. Bright \cii\ emission is associated
with bright CO emission and vice versa. The highest surface brightnesses
in the map are found towards N~11AB, containing LH~10 (\cii\ peak
surface brightness of $2.2 \times 10^{-4}\;\ergsr$).  This is also the
brightest \hii\ region, with half to two thirds of the H$\alpha$ and
thermal radio flux of the complex (cf. Table \ref{lmcclouds}).  Less
intense diffuse emission is present over a large part of the
area mapped. Both the intensity ($1.0\pm0.3 \times 10^{-4}\;\ergsr$)
measured by Boreiko \& Betz (1991) towards N11~B(CO) and their
upper limit for N11~B(H$_{2}$) are in rather good agreement with our
map results. Mochizuki \etal (1994) measured a beam-averaged intensity
$0.6\times 10^{-4}\;\ergsr$ in a 12.4$'$ beam, roughly corresponding
to the area mapped and shown in Figure 1.

Two \cii\ clouds are associated with the second-brightest \hii\ region
complex N~11C and N~11D (illuminated by LH~13 ) but the brightest of
the two \cii\ clouds (peak surface brightness $1.7\times
10^{-4}\;\ergsr$) occurs towards the weaker CO cloud.  There are other
significant differences.  Fig.~\ref{N11map} shows that the \cii\
emission extends far beyond the boundaries of the CO clouds, and that
it is largely continuous unlike the highly fragmented CO emission. The
contrast between \cii\ peaks and inter-cloud regions is much less in
\cii\ than in CO. \cii\ peaks are displaced from CO peaks in the
direction of the ionising OB stars. The shape of the extended \cii\
emission in Fig.~\ref{N11map} strongly suggests large-scale
dissociation of CO and subsequent ionisation of the resulting neutral
carbon.

The offset between peak CO and peak \cii\ emission is particularly
clear in N~11C, N~11D, and N~11E. The H$\alpha$ emission regions
N~11B, N~11C, N~11D, and N~11E are centered on the brightest stars of
their respective OB associations. These stars occur in the gaps
between \cii\ clouds 4 and 5, clouds 9 and 10, and to the southeast of
cloud 11, respectively. In all cases, the lines of sight towards the
stars and their HII regions show relatively little CO and \cii\
emission, which suggests that in those directions most carbon is
multiply ionised. The far-infrared emission from hot dust peaks at
these same positions consistent with the expected intense irradiation.
The association of emission from ionised carbon and hydrogen,
molecular gas, and warm dust is characteristic of PDRs (Kaufman \etal
1999).

\subsection{SMC}

\subsubsection{The N~66 Field}

%Figure 2  N66
\begin{figure}[]
\unitlength1cm
\begin{minipage}[t]{9cm}
\resizebox{8.57cm}{!}{\rotatebox{270}{\includegraphics*{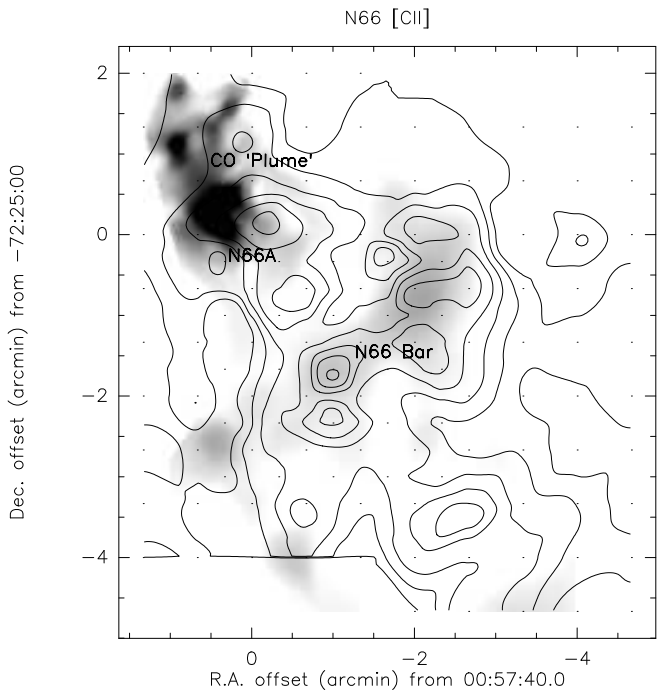}}}
\end{minipage}
\caption[]{Map of SMC-N~66. \cii\ contours are linear at multiples of
  $1.0\times10^{-5}\,\ergsr$.  Integrated $J$=2-1 $^{12}$CO emission
  is represented by grayscales in the range 0.5-5.0 $\kkms$. The
  locations of the N~66 bar, the bright \hii\ component N~66A, and the CO
  'plume' are indicated. Brightest \cii\ and CO emission occurs just
  off N~66A near position (0,0).}
\label{N66map}
\end{figure}

\begin{table*}
\scriptsize
\caption[]{SMC objects in [CII] maps }
\begin{center}
\begin{tabular}{lcccccccc}
\hline
\noalign{\smallskip}
Field & \multicolumn{6}{c}{Sources Included} 	    & \multicolumn{2}{c}{Integrated Flux} \\
      & Henize  & Davies & NGC & N/DEM Offset& IRAS & IRAS Position & F(H$_{\rm \alpha}$)  & S(843) \\ 
 &         & \etal &      & ($'$)      &           & $'$        & ($\ergs$)           & (mJy) \\
\noalign{\smallskip}     
\hline
\noalign{\smallskip}
1& N22     & DEM37  & 267 & --0.3, --1.6& LISMC 42  & --0.8, --1.8& 4.5 (-11); 10 (-11)&   95 \\
 & N25+N26 & DEM38  &     &  +0.1,  +0.5& LISMC 45  &  +0.1,  +0.6& 1.7 (-11); 6.7 (-11)&   40 \\
2& N19     & DEM31  &     & --3.5,  +0.5& anonymous & --3.8, --0.5& 2.5 (-11)           &  --- \\
 & ---     & DEM39  &     &     0, --4.6& --- 	    & --- 	  & ---	  	        &  --- \\
 & N27     & DEM40  &     &  +0.4, --0.3& LISMC 49  &  +0.6, --0.2& 1.5 (-11); 4.2 (-11)&   50 \\
3& N66	   & DEM103 & 346 & --0.7, --1.4& LISMC 131 & --1.0, --1.5&  80 (-11); 155 (-11)& 1640 \\
 & - N66-2 &        &     & --1.8, --1.5& ---	    & 	          & ---		        & 1420 \\
 & - N66-3 &	    &     & --0.8, --1.0& ---	    & 	          & ---		        &  120 \\
 & - N66-4 & 	    &     &  +0.4, --0.8& ---	    &	          & ---		        &  100 \\
4& N83A+C  & DEM147 & 456 & --0.7, --0.4&LISMC 199/200&--0.1, +0.2&  12 (-11); 18 (-11) &  290 \\
 & N83B    & DEM148 &     & --0.2,  +1.3& idem	    &             & included 	        & included \\
 & N84C    & DEM149 &     &  +1.6,  +1.3& idem	    & 	          & 0.5 (-11); 2.6 (-11)&   12 \\
 & 	   & DEM150 &     &  +2.1,  +3.1& anonymous &  +2.3,  +3.2& 2.0 (-11) 	        &  --- \\
 & N84A    & DEM151 & 460 &  +3.2, --0.8& LISMC 201 &  +3.5, --0.7&  11 (-11); 14 (-11) &  240 \\
 & N84B+D  & DEM152 &     &  +4.4, --2.6& LISMC 202 &  +3.8, --3.6& 6.8 (-11) 	        & included \\
 & ---     & DEM154 &     &  +5.5, --3.1& LISMC 203 &  +6.1, --1.7&	---	  	&      \\
5& N88	   & DEM161 &     &     0, --0.4& LISMC 215 & --0.1, --0.3& 4.0 (-11); 4.1 (-11)&  210 \\
\noalign{\smallskip}     
\hline
\end{tabular}
\end{center}
%Notes to Table 1:\\
\label{smcclouds}
\end{table*}

\begin{table*}
\scriptsize
\caption[]{SMC [CII] data}
\begin{center}
\begin{tabular}{lcccccccc}
\noalign{\smallskip}     
\hline
\noalign{\smallskip}
Object 	& Peak Intensity & Position   & \multicolumn{3}{c}{Integrated Flux} & \multicolumn{3}{c}{Ratios}   \\
  & $I_{\rm CII}$ & Offset & $F_{\rm CII}$ &$F_{\rm CO}$ & $F_{\rm IR}$
& {$F_{\rm CII}\over F_{\rm CO}$} & {$F_{\rm CII}\over F_{\rm IR}$} &
{$F_{\rm IR}\over F_{\rm CO}$} \\
 & ($\ergsr$) & ($''$)& \multicolumn{3}{c}{($\ergs$)} & & & \\
\noalign{\smallskip}
\hline
\noalign{\smallskip}
N22+N25+N26& 7.3 (-5)&     0, 0    & 2.3 (-11) & 1.6 (-15)& 3.4 (-9) & 1.5 (4)& 0.7 (-2) & 2.3 (6) \\
N27    	   & 12  (-5)&  +0.6, --0.1& 5.1 (-11) & 2.7 (-15)& 3.6 (-9) & 1.9 (4)& 1.4 (-2) & 1.3 (6) \\
N19  	   & 5.5 (-5)& --3.3, --0.7& 2.3 (-11):&  ---     & 1.5 (-9):&  ---   & 1.5 (-2):&   ---   \\
N27-S  	   & 3.4 (-5)&  +0.7, --2.7& 1.4 (-11) &  ---     & 1.7 (-9):&  ---   & 0.8 (-2):&   ---   \\
DEM39      & 3.0 (-5)&  +0.7, --4.0& 0.7 (-11):&  ---     & 0.4 (-9):&  ---   & 1.7 (-2):&   ---   \\
N66    	   & 6.9 (-5)& --1.0, --1.7& 6.4 (-11) & 1.0 (-15)& 9.6 (-9) & 6.4 (4)& 0.7 (-2) & 9.6 (6) \\
N83ABC+N84C & 6.7 (-5)&  +0.2,  +0.3& 4.8 (-11) & 1.4 (-15)& 3.0 (-9) & 3.4 (4)& 1.6 (-2) & 2.1 (6) \\
DEM150     & 1.9 (-5)&  +2.3,  +3.2& 0.7 (-11):&  ---     & 0.4 (-9)&  ---    & 1.7 (-2):&   ---   \\
N84-N	   & 2.8 (-5)&  +4.0,  +3.3& 1.7 (-11) &  ---     & 1.1 (-9):& ---    & 1.6 (-2):&   ---   \\
N84  	   & 6.2 (-5)&  +3.3, --1.3& 2.0 (-11) &  ---     & 2.2 (-9) & ---    & 0.9 (-2) &   ---   \\
- N84A	   & 6.2 (-5)&  +3.3, --1.3& 0.5 (-11) &  ---     & 0.7 (-9):& ---    & 0.7 (-2):&   ---   \\
- N84B 	   & 6.0 (-5)&  +4.0, --3.0& 1.5 (-11) & 0.2 (-15)& 1.6 (-9):& 8.3 (4)& 0.9 (-2):& 8.9 (6) \\
DEM154 &$\leq$0.7 (-5)& +5.5, --3.1& ---       &  --- & $\leq$0.4 (-9)& ---   & ---      &   ---   \\
N88    	   & 6.3 (-5)&     0, 0    & 1.1 (-11) & 0.1 (-15)& 2.4 (-9) & 8.2 (4)& 0.5 (-2) & 18 (6)  \\
\noalign{\smallskip}     
\hline
\end{tabular}
\end{center}
%Notes to Table 1:\\
\label{smcdat}
\end{table*}

N~66, excited by the OB association NGC~346, is the largest and
brightest \hii\ region in the SMC.  It is located about midway in the
main body of the SMC Bar and extends over an area of $3 '\times 6'$
(linear size $50\times 110$ pc). NGC~346 is a young cluster of age
about 3 Myr, and it contains several O stars among which are O3 -O5
stars (Walborn \& Blades 1986; Evans \etal 2006 and references
therein). Star formation has occurred in this region for much longer,
up to 10 Myr, and must be ongoing as N~66 contains numerous
pre-main-sequence objects (Sabbi \etal 2007; Simon \etal 2007;
Hennekemper \etal 2008; Gouliermis \etal 2006, 2008, 2009) as well as
a supernova remnant (Ye \etal 1991).  Although the nebula shows
considerable structure, there are no strong peaks in either line or
continuum emission from gas and dust.  The \hii\ region proper
consists of an extended envelope of diffuse ionised gas in which a
bright bar-like structure (southeast to northwest) is embedded.  A
bright region (N~66A) occurs just off the Bar to the northeast. Much
of the mid-infrared line and continuum emission exhibit the same
structure (Rubio \etal 2000; Contursi \etal 2000), as indeed does the
distribution of the bright stars themselves (cf. Gouliermis \etal
2006; Hennekemper \etal 2008) as illustrated very nicely by public HST
images. Recent star formation along the N~66 Bar may have been
triggered by expanding windblown shells (Gouliermis \etal 2008).

Rubio \etal (2000) mapped carbon monoxide in N~66 in the $J$=1-0 and
$J$=2-1 transitions at resolutions of $43''$ and $21''$, respectively.
The CO maps resemble the general picture seen at other wavelengths.
Much of the CO follows the N~66 Bar (as does hot $\h2$ also observed
by Rubio \etal 2000), but much stronger CO emission occurs in a
linear `plume' pointing northeast at right angles to the N~66 Bar. It
almost appears as if N~66A is the `burning' end of the feature. The
distribution of \cii\ in N~66 closely resembles the H$\alpha$ and CO
distributions shown by Ye \etal (1991) and Rubio \etal (2000). The
lowest contour in our map (Fig.~\ref{N66map}) roughly follows the
contour of 15--20\% of the H$\alpha$ peak brightness.  The highest
\cii\ intensities occur in the N~66 Bar where the presence of
near-infrared $K$-band $\h2$ emission suggests strong irradiation of
the ambient ISM.  A secondary \cii\ maximum is to the northeast of the
bar, almost precisely adjacent to the strong CO `plume'.  This
\cii\ maximum is relatively weak with respect to the adjacent CO, but
relatively strong with respect to the almost coincident H$\alpha$ peak
(cf. Fig. 3 by Ye \etal 1991). It suggests a PDR seen from the side.
Apart from scale, N~66 is similar to the LMC \hii\ region complexes 30
Doradus and N~160 which combine relatively small CO clouds with rather
extended \cii\ emission (Johansson \etal 1998; Poglitsch \etal 1995;
Israel \etal 1996). Like these sources, N~66 appears to be a
relatively evolved object in which most of the original molecular gas
has been consumed by star formation or processed by irradiation. Much
of the remaining CO has been photo-dissociated, and much of the
resulting C$^{o}$ has been turned into the extended cloud of C$^{+}$
shown in Fig.~\ref{N66map}.  Clearly, this process is still actively
going on near N~66A.

\subsubsection{The N~83/N~84 Field}

%Figure 3  N83/N84
\begin{figure}[]
\unitlength1cm
\begin{minipage}[t]{9cm}
\resizebox{8.57cm}{!}{\rotatebox{270}{\includegraphics*{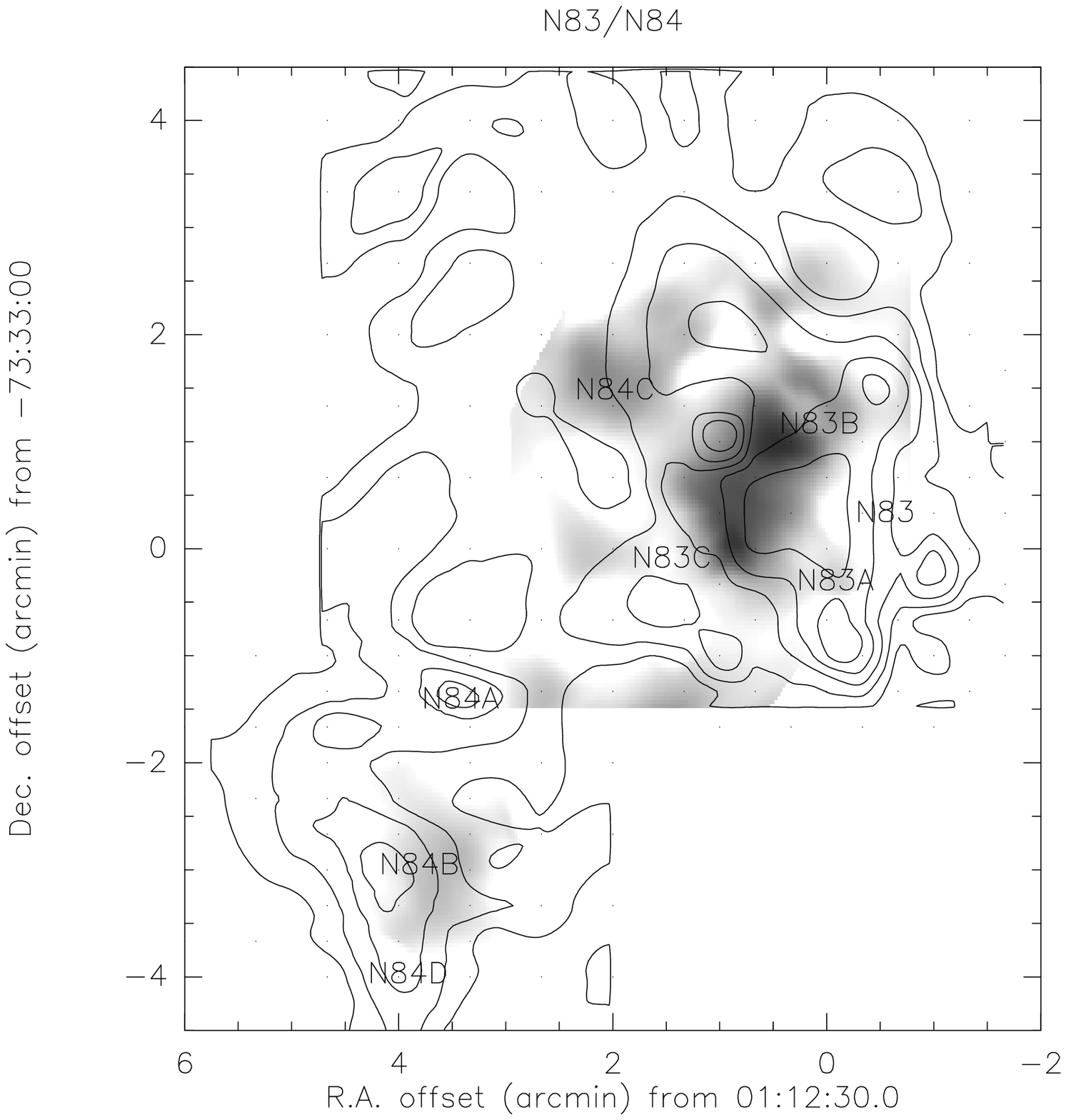}}}
\end{minipage}
\caption[]{Map of SMC-N~83/N~84/N~85. \cii\ contours are linear at
  multiples of $1.0\times10^{-5}\,\ergsr$.  Integrated $J$=1-0
  $^{12}$CO emission is represented by grayscales in the range 0.5-4.5
  $\kkms$. The positions of the (mostly compact) \hii\ regions in the
  complex are marked.}
\label{N83map}
\end{figure}

The SMC Wing has a much lower stellar and gaseous surface density than
the SMC Bar, but nevertheless contains a number of mostly compact but
intense \hii\ regions.  Most prominent is the N~83/N~84/N~85 complex
(DEM~147-152) which consists of several small nebulae each excited by
a single star or a tight group of stars (Testor \& Lortet 1987) in the
NGC~456 and NGC~460 associations. The spectral type of the ionising
stars ranges from O4 to O9, the associated \hii\ mass is only $\sim
5000$ M$_{\sun}$, and typical visual extinctions are between 0.5 and
1.0 magnitude (Copetti 1990; Caplan \etal 1996).  Relatively bright
far-IR emission coincides with the complex (LI-SMC 199-202; Schwering
\& Israel 1990). SEST CO observations of this field have been
discussed by Israel \etal (2003b), Bolatto \etal (2003), and Leroy
\etal (2009). Fig.~\ref{N83map} shows contours of the widespread
\cii\ emission and the CO map from Israel \etal (2003b) in shades of
grey. For a map of the CO emission superposed on a red optical image,
identifying the various \hii\ regions in more detail, we refer to
Bolatto \etal (2003).

Again, we find \cii\ and CO to be loosely correlated. There is an
extended \cii\ cloud (dimensions of $2.6'\times4.6'$, i.e.
$45\times80$ pc; peak surface brightness about $7\times10^{-5}$
\ergsr) in the same direction as the large nebula N~83, suspected by
Bolatto \etal (2003) to be a supernova remnant. The brightest CO cloud
appears to embrace the northeastern boundary of N~83 and contains the
compact objects N~83B and N~83C.  The \cii\ emission is mostly {\it
  adjacent to} the CO cloud and coincident with N~83 and its brightest
part, N~83A, and wraps around the CO cloud in the north. The strong
infrared counterpart, LI-SMC 199/200, is offset to the north by 14 pc
from the \cii\ peak.

The small \hii\ region N~84C is associated with a distinct but minor
CO cloud. There is no corresponding \cii\ peak; rather it is on a
\cii\ gradient.  N~84A is an extended \hii\ region on the surface of a
large shell (55 pc) of CO emission. Much of the complex is at a
{\it minimum} in \cii\ emission.  The infrared counterpart LI-SMC 201 is 17
pc north of the \hii\ region N~84A and its associated \cii\ peak.
This \cii\ peak is part of a ridge of ionised carbon (peak surface
brightness about $6\times10^{-5}$ \ergsr) that starts at the eastern
edge of the CO shell and extends southwards to the compact
\hii\ regions N~84B and N~84D in the lower left corner of
Fig.~\ref{N83map}. The \cii\ maximum associated with N~84 has an
elongated shape of about $1.3' \times 3.5'$ ($23 \times 61$ pc).  The
compact objects N~84~B and N84~C are embedded in a distinct but not
very bright CO cloud. As the map shows, the \cii\ emission is mostly
adjacent to this CO cloud.  It overlaps with the extended H$\alpha$
emission seen in optical images but its maximum is just south of the
ionised gas and about 8 pc east of the CO cloud. The infrared peak
LI-SMC 202 occurs between the \cii\ emission peaks from N~84A and
N~84B and is about 12 pc north from the CO peak. Finally, there is an
extended region of diffuse \cii\ emission with a surface brightness of
$2\times10^{-5}$ \ergsr\ ($2\sigma$) covering the area east of N~83
and north of N~84 (upper right in Fig.~\ref{N83map}).  It has only a
very weak CO counterpart.

\subsubsection{Other fields: N~22/N~25/N~26, N~19/N~27, and N~88}

%Figure 4  N25+N88
\begin{figure*}[]
\unitlength1cm
\begin{minipage}[t]{4.0cm}
\resizebox{4.2cm}{!}{\rotatebox{270}{\includegraphics*{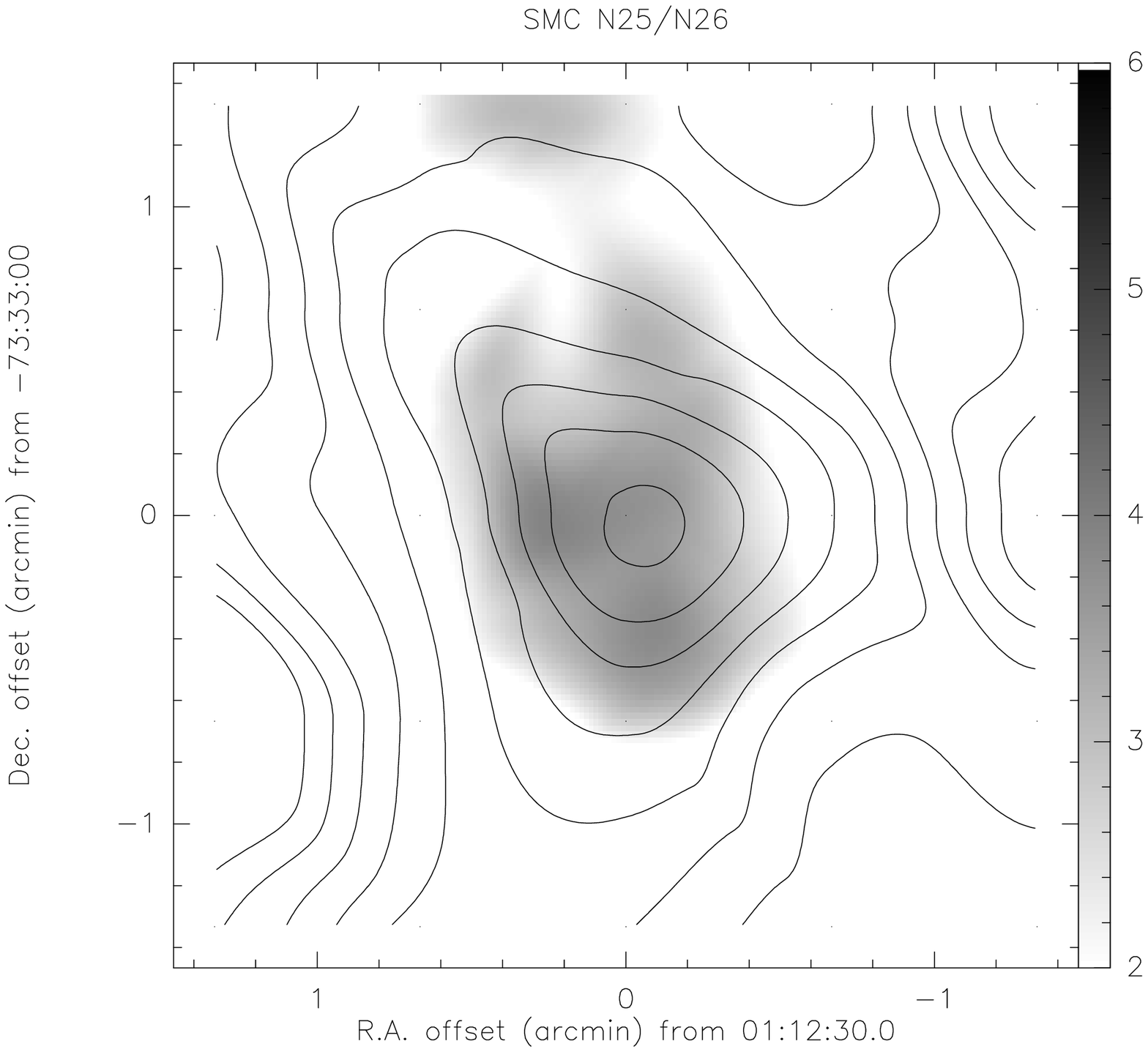}}}
\end{minipage}
\hfill
\begin{minipage}[t]{10cm}
\resizebox{10cm}{!}{\rotatebox{270}{\includegraphics*{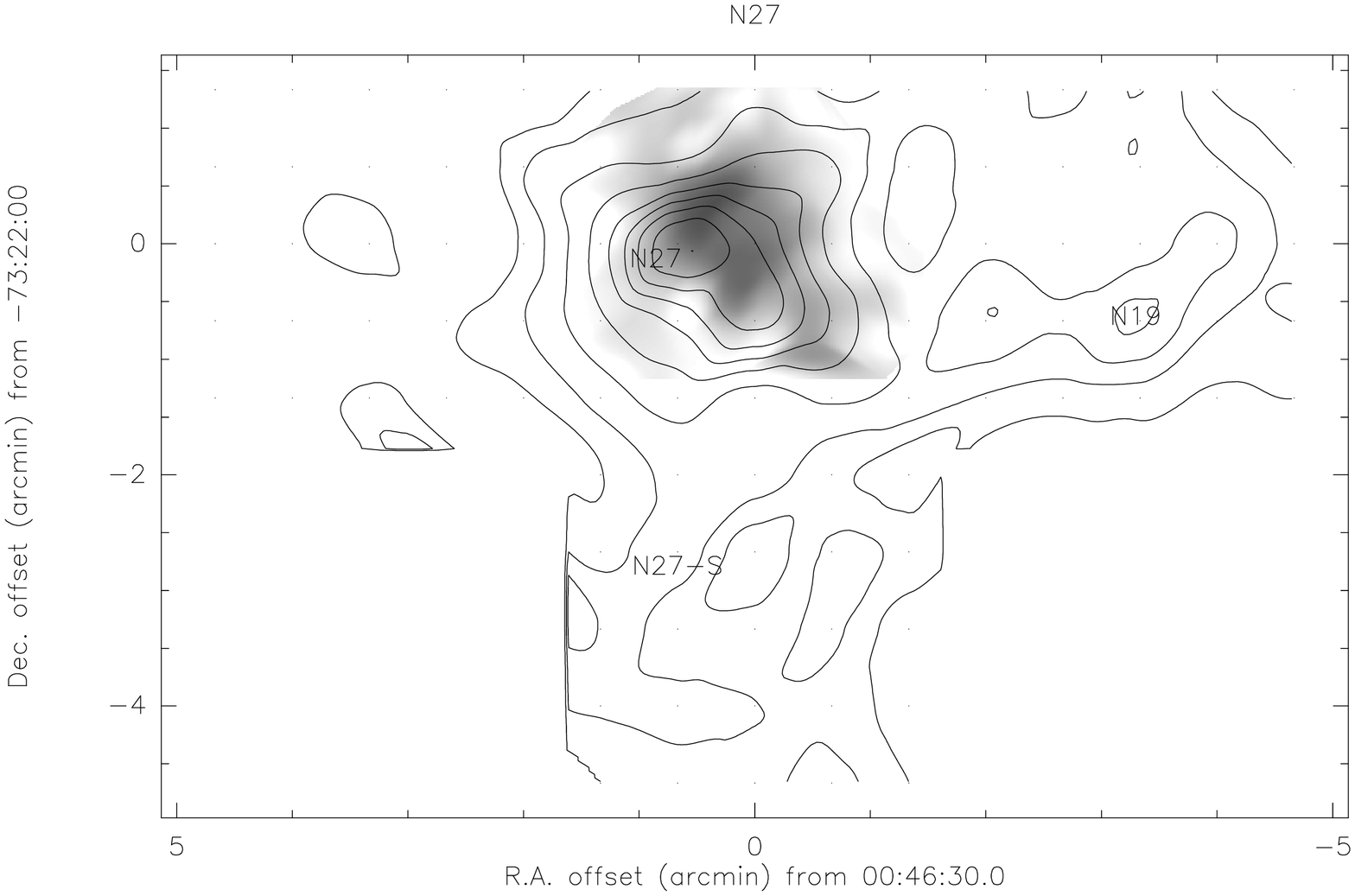}}}
\end{minipage}
\hfill
\begin{minipage}[t]{4.0cm}
\resizebox{4.3cm}{!}{\rotatebox{270}{\includegraphics*{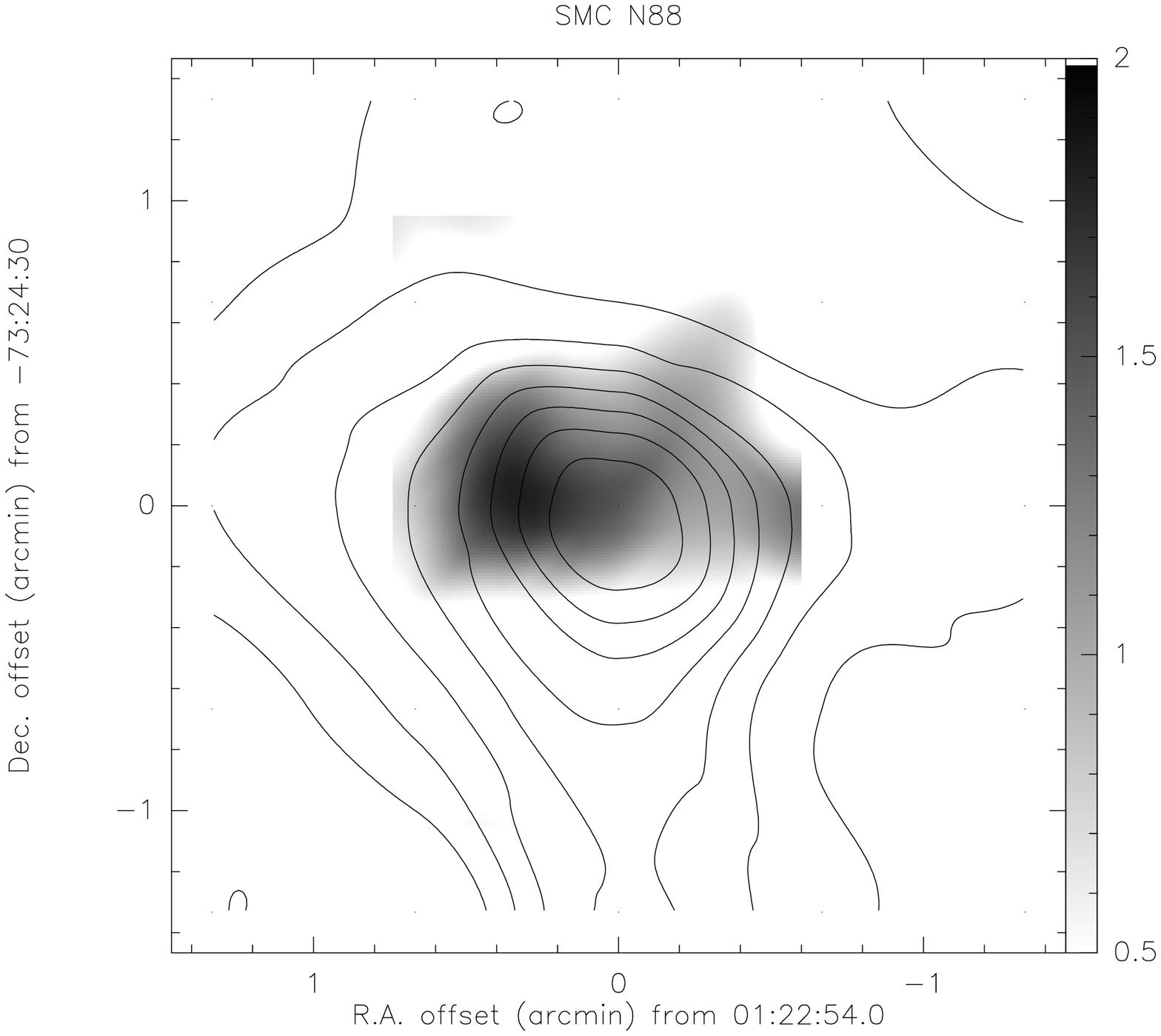}}}
\end{minipage}
\caption[]{Left: \cii\ Map of SMC-N~25 field. Contours are linear at
  multiples of $0.72\times10^{-5}\,\ergsr$.  Integrated $J$=1-0
  $^{12}$CO emission is represented by grayscales in the range 0.5-5.0
  $\kkms$. Center: \cii\ Map of SMC-N~27 field. Contours are linear at
  multiples of $1.35\times10^{-5}\,\ergsr$.  The positions of the
  \hii\ regions N27, N27-S, and N~19 are indicated. Integrated
  $J$=1-0 $^{12}$CO emission is represented by grayscales in the range
  1.0-13.0 $\kkms$. Right: \cii\ map of SMC-N~88 field. Contours are
  linear at multiples of $0.72\times10^{-5}\,\ergsr$.  Integrated
  $J$=1-0 $^{12}$CO emission is represented by grayscales in the range
  0.2-2.4 $\kkms$. }
\label{N27map}
\end{figure*}

\nobreak {\it The N~22/N~25/N~26 field} (Fig.~\ref{N27map}, left)
contains a single \cii\ source of diameter $1.3 ' \times 2.1 '$ ($23
\times 36$ pc), located between the \hii\ regions N~25/N~26 and N~22
(NGC 267), but closest to the former. The largest \hii\ region N~22
(DEM 37: $2'$) is located $98''$ (28 pc) south from the \cii\ peak,
marking the southern edge of the \cii\ source.  The brighter and more
compact \hii\ regions N~25 and N~26 (DEM 38: $1'$) are located $33''$
(10 pc) north of the [CII] peak (see Fig. 1 by Testor 2001) and mark the
northern edge of the source.  Both N~25/N~26 and N~22 are relatively
strong 843 MHz radio continuum sources (Mills \& Turtle 1984) in the
populated southern half of the SMC Bar.  N~25 and N~26 coincide with
the infrared source LI-SMC 45. Diffuse \cii\ emission, at a typical
level of $2.7 \times 10^{-5}$ \ergsr\ (i.e. $3\sigma$), extends beyond
the map boundaries.  All three \hii\ regions appear to be excited by
single O-stars (Hutching \& Thompson 1988; Testor, 2001). As
Fig.~\ref{N27map} illustrates, the \cii\ emission is very nearly
coincident with the CO cloud (SMC-B2 no. 3) mapped by Rubio \etal
(1993) just south of N~25 and N~26. The \hii\ region N~22 appears to
fill a hole in the CO distribution.  Thus, the observed \cii\ emission
closely follows the CO distribution, and its major source of
excitation are N~25 and N~26 (LI-SMC 45) at a nominal projected
distance of $34''$ or 9.5 pc.

\nobreak {\it The N~19/N~27 field} (Fig.~\ref{N27map}, center)
contains a bright \cii\ source of diameter $1.9'$ (33 pc) offset to
the north by about $5''$ from the compact \hii\ region N~27 (DEM 40:
$0.8'$). N~27 has a strong counterpart in the 843 MHz radio continuum
map by Mills \& Turtle (1984).  It is located at the southern edge of
a relatively bright and compact CO cloud of diameter $1.4'$
(deconvolved linear size of 21 pc).  As Fig.~\ref{N27map} shows, the
CO peak is offset from the \cii\ source by $25''$ to the
northwest. The higher-resolution $J$=2--1 \twelveco\ map published by
Rubio \etal (1996) shows the bulk of the CO emission to be in a curved
ridge actually adjacent to the \cii\ peak (projected peak-to-peak
separation 6 pc). The strong infrared source LI-SMC 49 is very close
to the \cii\ peak. Thus, both FIR and [CII] peak in between the
\hii\ region and the CO cloud, at about 5 pc projected distance from
either.
 
At a level of $3.3 \times 10^{-5} \ergsr$ (i.e. $4\sigma$), diffuse
emission extends to the west. One of the minor maxima in this emission
appears to be associated with the diffuse \hii\ region N~19 (DEM 31:
1.8$'$). The eastern boundary of the \cii\ emission is sharply
defined, but weak diffuse \cii\ emission also extends southwards
towards the compact \hii\ region DEM 39 (0.5$'$) which is off the
map. To the east and north of DEM 39, relatively weak CO clouds are
found (see Fig. 3 in Rubio \etal 1993). Further comparison shows that
the \cii\ emission appears to skirt the radio emission from a complex
of sources southwest of N~27, containing at least one extended
supernova remnant.

\nobreak {\it The N~88 field} (Fig.~\ref{N27map}, right) contains a
single \cii\ source with a deconvolved diameter of $1'$ ( 17 pc). The
very bright and compact HII region N~88A/N~88B (Testor \& Pakull 1985;
Testor \etal 2010) are located $23''$ (7 pc) south of the \cii\ peak
and coincides with the infrared source LI-SMC 215. `The \twelveco\ and
\cii\ peaks almost coincide (see Fig.~\ref{N27map}), but the CO
extends to the north, whereas the \cii\ has an extension to the south,
encompassing the HII region.  In the compilation by Caplan
\etal\ (1996), N88 has the largest extinction, $A_{\alpha}'$ = 1.34
mag, of all three dozen \hii\ regions listed (mean value 0.50 mag). As
in previous cases, the \cii\ emission extends well beyond the CO
boundaries.

\subsection{Fluxes and Luminosities}

\begin{table*}
\scriptsize
\caption[]{PDR Parameters, Masses, and Luminosities  }
\begin{flushleft}
\begin{tabular}{lcccccccccc}
\hline
\noalign{\smallskip}
 Object & $A_{\rm eff}^{a}$ & $A_{\rm eff}^{b}$ & \multicolumn{2}{c}{$G_o$} 
 & $L_{\rm CII}$ & $L_{\rm IR}$ & $L_{\rm CO}^{c}$  & $N_{\rm H}(min)^{d}$ 
 & $M_{\rm H}$(min) & $\mvt$(CO)$^{e}$ \\ 
 & ($10^{-7}\,$sr)&($10^{40}\,$cm$^2$)&L$_{\rm Lyc}^{f}$
 & $I_{\rm IR}^{g}$ &($10^3\,$\lsol)&($10^5\,$\lsol)&($10^{3}\,\kkms\,\pc2$)
 & ($10^{21}\,$cm$^{-2}$)&\multicolumn{2}{c}{$10^4\,$\msol)} \\   
\noalign{\smallskip}     
\hline
\noalign{\smallskip}
\multicolumn{11}{c}{LMC}\\
\noalign{\smallskip}     
\hline
\noalign{\smallskip}
N~11 (1)     & 4.0 & 0.95 & --- & 17 &  2.3  &  0.7  & 1.8 & 0.6 & 0.6 & 1.5 \\ 
N~11F (2+3)  & 4.5 & 1.1  &  45 & 41 &  3.2  &  1.8  & 1.1 & 0.8 & 0.9 & 1.1 \\
N~11AB (4+5) & 6.2 & 1.5  & 178 &180 & 11    & 11    & 3.6 & 1.8 & 2.9 & 2.7 \\ 
N~11 (6)     & 2.3 & 0.55 & --- &--- &  2.3  & ---   & 0.9 & 1.1 & 0.7 & 0.9 \\
N~11 (7)     & 2.7 & 0.65 & --- &--- &  0.05 & ---   & 1.2 & 0.2 & 0.2 & 1.1 \\
N~11 (8)     & 2.8 & 0.65 & --- &--- &  1.4  & ---   & 0.7 & 0.5 & 0.4 & 0.8 \\
N~11C (9)    & 4.5 & 1.1  &  25 & 27 &  5.9  &  2.0  & 0.7 & 0.8 & 1.6 & 0.7 \\
N~11D (10)   & 4.2 & 1.0  &  25 & 27 &  3.9  &  2.0  & 3.6 & 0.6 & 1.1 & 2.8 \\
N~11E (15)   & 5.8 & 1.4  &  22 & 35 &  3.2  &  1.4  & 4.2 & 0.8 & 0.8 & 3.1 \\
\noalign{\smallskip}     
\hline
\noalign{\smallskip}
\multicolumn{11}{c}{SMC}\\
\noalign{\smallskip}     
\hline
\noalign{\smallskip}
N~22+N~25+N~26&3.2 & 1.1  & 138 & 62 &  2.6  &  3.8  & 3.6 & 3.1 & 3.7 & 2.7 \\
N~27         & 4.3 & 1.5  &  40 & 62 &  5.7  &  4.1  & 6.1 & 5.1 & 8.4 & 4.2 \\
N~19  	     & 4.2 & 1.4  &  30 & 23 &  2.6  &  1.7  & --- & 2.4 & 3.7 & --- \\
N~27-S       & 4.1 & 1.4  & --- & 12 &  1.6  &  1.9  & --- & 1.5 & 2.3 & --- \\
DEM~39       & 2.3 & 0.8  & --- & 12 &  0.8  &  0.5  & --- & 1.3 & 1.1 & --- \\
N~66         & 9.3 & 3.2  & 954 &127 &  7.2  & 11.0  & 2.3 & 3.0 &10.8 & 1.9 \\
N~83AB+N~84C & 7.2 & 2.5  & 298 & 50 &  5.4  &  3.4  & 3.3 & 2.9 & 7.9 & 2.5 \\
DEM~150      & 3.7 & 1.3  & --- & 15 &  0.8  &  0.5  & --- & 8.1 & 1.1 & --- \\
N~84-N	     & 6.1 & 2.1  & --- &  6 &  1.9  &  1.2  & --- & 1.2 & 2.8 & --- \\
N~84  	     & 3.2 & 1.1  & 208 & 27 &  2.3  &  2.5  & --- & 2.7 & 3.2 & --- \\
- N~84A	     & 0.8 & 0.3  & 139 & 35 &  0.6  &  0.8  & --- & 2.7 & 0.8 & --- \\
- N~84B	     & 2.5 & 0.85 &  69 & 46 &  1.7  &  1.8  & 0.4 & 2.6 & 2.5 & 0.5 \\
N~88         & 3.2 & 0.6  &  74 & 36 &  1.2  &  2.7  & 0.2 & 2.7 & 1.7 & 0.3 \\
\noalign{\smallskip}     
\hline
\end{tabular}
\end{flushleft}
%Notes to Tables 6 and 7\\
$^{a}$ Calculated as the ratio of the integrated flux to
the peak intensity 
$^{b}$ All size scales, luminosities and masses assume 
distances to the LMC $D=50$ kpc and SMC $D=60$ kpc
$^{c}$ The CO luminosity is in units of $10^3\;{\rm K\;
km\;s^{-1}\;pc^2}$
$^{d}$ Calculated in the high-density, high-temperature
limit; see text. 
$^{e}$ 'Virial' molecular mass calculated from the CO luminosity as
$\mvt$(CO) = $39 L_{\rm CO}^{0.8}$. 
$^{f}$ $G_o$ calculated from the estimated Lyman continuum
luminosity and a mean distance from the ionising source to the cloud
derived from the projected extent of the emission;
$L_{\rm Lyc}$ derived from either H$\alpha$ measurements (ref) or 843
MHz radio observations (ref) with an assumed black-body effective
temperature of $T_{\rm BB}=50,000$ K.
$^{g}$ $G_o$ estimated from the far-infrared surface
brightness, using $G_o=I_{\rm IR}/1.3\times 10^{-4}\;\ergsr$.
\label{PDR}
\end{table*}

\begin{figure*}[]
\unitlength1cm
\begin{minipage}[t]{18cm}
\resizebox{6.4cm}{!}{\rotatebox{0}{\includegraphics*{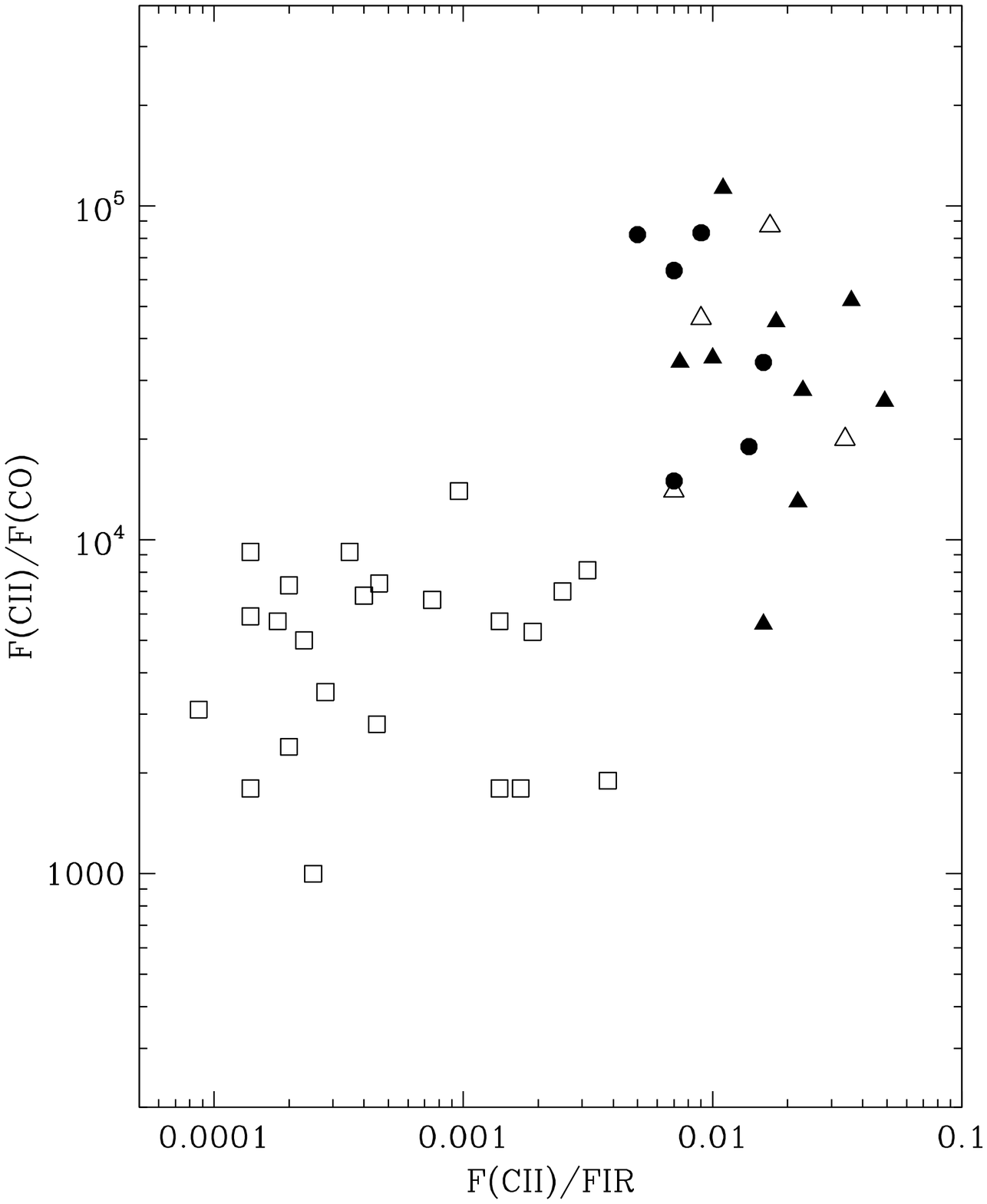}}}
\resizebox{6.4cm}{!}{\rotatebox{0}{\includegraphics*{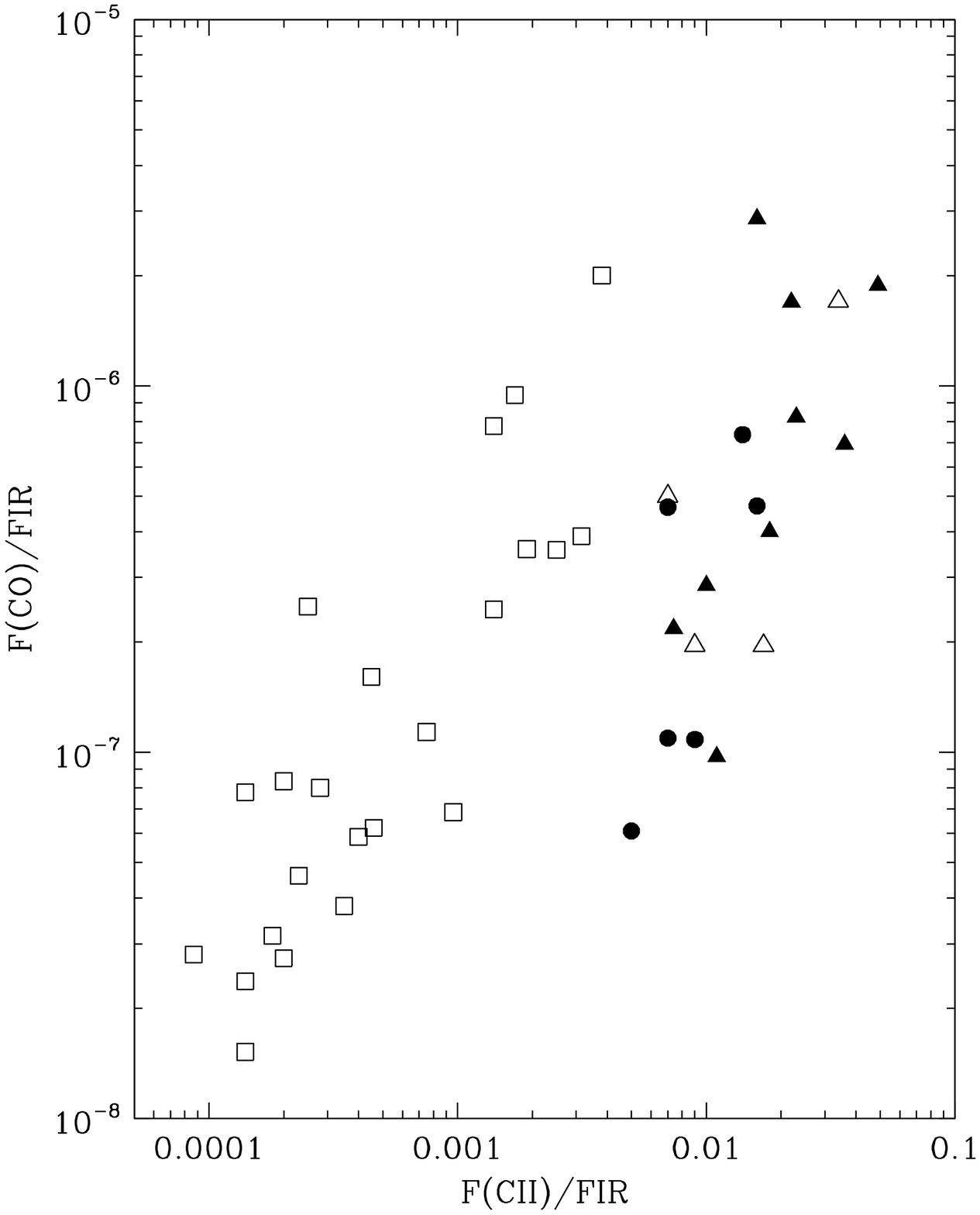}}}
\resizebox{6.4cm}{!}{\rotatebox{0}{\includegraphics*{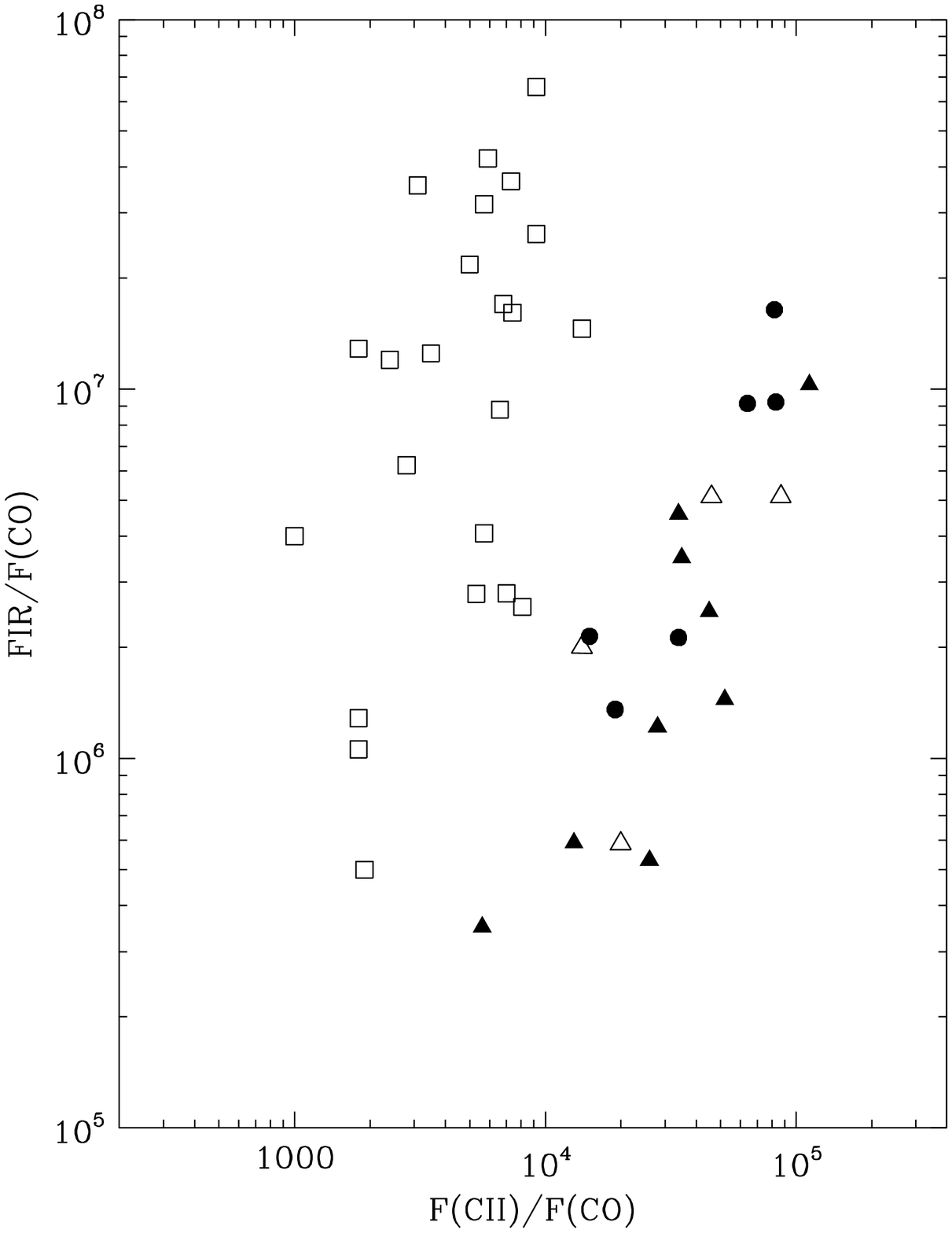}}}
\end{minipage}
\caption[]{Diagrams illustrating the various ratios of [CII], CO and
  FIR emission for sources in the LMC (filled triangles, data from
  this paper, Israel \etal 1998, and Poglitsch \etal 1995), the SMC
  (filled circles, data from this paper), IC~10 (open triangles, data
  from Madden \etal. 1997), and the Milky Way (open squares, data from
  Table 5 in Stacey \etal 1991). The shift of sources in the
  low-metallicity galaxies LMC, SMC, and IC~10 with respect to those
  in the Milky Way is very clear and due to relatively brighter [CII]
  emission caused by greater UV mean free path lengths in the low
  metallicities of the Magellanic Clouds and IC~10. One might expect
  the low-metallicity SMC sources to be shifted even more than the
  intermediate-metallicity LMC sources, but the opposite is true. The
  lesser SMC shift reflects lower [CII] photo-electric heating due to
  the depletion of PAHs in that galaxy. See Sect. 4 for further
  details.  }
\label{ratio}
\end{figure*}

\nobreak All \cii\ results are summarised in Tables~\ref{lmcdat} and
\ref{smcdat}, along with the relevant infrared and CO data. Peak
intensities and positions with respect to the reference position from
Table~\ref{kaofields} are given in Columns 2 and 3; the integrated
\cii\ intensity is listed in column 4. Two SMC \hii\ regions (DEM~39,
DEM~150) do not coincide with a peak in the \cii\ emission: here we
have taken the actual \cii\ intensity at the position listed; their
estimated integrated \cii\ intensities are uncertain. No
\cii\ emission was detected towards DEM~154; we calculate an upper
limit to the luminosity assuming a source size of $1'$. In the next
section we make quantitative estimates of the physical conditions
characterising the emitting regions; however, some important
inferences can be drawn simply from examination of the data given in
Tables 3 and 5. The various ratios of [CII], CO and far-infrared
emission listed in those Tables are illustrated in
Fig.\,\ref{ratio}. For comparison, we have also included the relevant
ratios for sources in the Milky Way (taken from Stacey \etal 1991) and
in the dwarf irregular galaxy IC~10 (Madden \etal 1997).

%Figure 5  Ratios

The peak intensities observed in the 158\mm\ line in the N~11 clouds
are typically $I_{\rm CII}\approx 1-2\,\times 10^{-4}\;\ergsr$, very
similar to the values measured elsewhere in the LMC by Poglitsch \etal
(1995) and Israel \etal (1996): in 30 Doradus, in N~159, and in
N~160. In contrast, the peak intensities of the SMC clouds are lower
by a factor of $\sim 3$.  Only the highest measured value, that of
N~27, just exceeds $10^{-4}\;\ergsr$.  N~27 (LI-SMC~49, also known as
LIRS-49) is an unusually bright object for the SMC. It is also the
second brightest CO cloud in the SMC.  The integrated fluxes in the
158 \mm\ line and the infrared continuum \footnote{As in Paper I, the
  far-infrared fluxes are given as the {\sl IRAS} FIR parameter, a
  weighted sum of the 60 and 100\mm\ fluxes: FIR = (1.26F(60) +
  F(100)).}  are smaller than those in the observed LMC regions by a
similar factor. The {\it ratios} of the 158\mm\ line to far-infrared
continuum fluxes for the SMC sources are about half those seen in the
LMC, at $0.7-1.7\%$, but still well above those of comparable Galactic
sources. Similarly high values are seen in the nearby, low-metallicity
irregular galaxy IC 10 (Madden \etal 1997). As we discuss below, this
ratio has important implications for the efficiency of grain
photoelectric heating and the typical FUV radiation field
characterising PDRs in the SMC.

One of the characteristics that has also emerged from the study of
$\cp$ 158 \mm\ line emission from low-metallicity galaxies (Poglitsch
\etal 1995; Israel \etal 1996; Madden \etal 1997) is a high ratio of
the 158\mm\ line flux to that of \twelveco\ $J=1-0$ line. The typical
value seen in ``warm dust'' galactic nuclei or Galactic \hii\ regions
is $\fcii/\fco\approx 4000-6000$, while in ``cool dust'' galaxies and
Galactic giant molecular clouds (GMCs) the ratio is approximately 1500
(Stacey \etal 1991). In contrast, the ratio observed in the
infrared-luminous regions of the N159 and N160 complex range from 5600
to $34,000$ (Israel \etal 1996), while it is approximately $10^5$ in
the massive star-forming complex 30 Doradus (Poglitsch \etal
1995). Table~\ref{lmcdat} gives equally high ratios
$\fcii/\fco\approx 20,000-260,00$ for the clouds in N~11. In the SMC
regions, where we have relevant CO intensities for about half the $\cp$
sample, we calculate similar ratios, ranging from $15,000$ to $83,000$.
Three out of the six regions with CO data exhibit ratios $\fcii/\fco
\ge 60,000$. As we will discuss in \S 4.1, such large ratios are a
direct consequence of the reduced metallicity and dust-to-gas ratios
in the Magellanic clouds.

We also note that most of the regions observed in the LMC have
$\fir/\fco$\ ratios comparable to Galactic GMCs or IR-bright galactic
nuclei, but substantially below the value seen in Galactic
\hii\ regions (see the discussion in Paper I). Only 30 Doradus
approaches the typical \hii\ region value. In contrast, half of the
SMC sample (the regions with the highest $\fcii/\fco$ ratios) show
$\fir/\fco$ ratios similar to the Galactic \hii\ region value.
However, as we discuss below, the similarity of this ratio is
misleading, because the physical conditions in the SMC regions are
markedly different from those in Galactic \hii\ regions (as evidenced
by the order-of-magnitude larger values of both $\fcii/\fco$ and
$\fcii/\fir$ in the SMC regions).

\section{Physical Conditions and PDR Parameters}
\nobreak As first demonstrated by Crawford \etal (1985) and Stacey
\etal (1991), much of the 158\mm\ line emission of $\cp$ originates in
PDRs: the interfaces between dense molecular clouds and the \hii\
regions surrounding luminous, early-type stars. Theoretical studies of
high-density, high-UV flux PDRs were pioneered by Tielens \&
Hollenbach (1985a,b). A summary of both theory and observation is
given in Hollenbach \& Tielens (1997).  PDRs are luminous sources of
158\mm\ line emission because there is an extensive zone in which
carbon is kept largely in the form of $\cp$ through ionisation by FUV
photons, and the gas temperature is kept high enough (relative to the
line energy above ground of 91 K) that the 158\mm\ transition is
easily excited. The extent of the $\cp$ zone is determined by
attenuation of the FUV photons by dust. As we will show below, this
leads to a crucial scaling: if the dust-to-gas ratio and the gas-phase
carbon abundance are varied by the same factor, the column density of
singly-ionised carbon will remain essentially constant. This has
profound implications for the nature of PDRs and the interstellar
medium in low-metallicity galaxies.

In all cases we are assuming that all of the $\cp$ emission arises
from PDRs, and we have ignored any contribution from the \hii\ regions
themselves. Although this is not strictly correct, it is unlikely that
this assumption has any significant impact on the analysis, for two
reasons. (1) In Galactic star-forming complexes, the contribution from
the \hii\ regions to the total $\cp$ emission is minor, $\sim 10\%$ or
less (e.g., Stacey et al 1991), since much of the carbon is more
highly ionised than $\cp$.  Theoretical models (Kaufman, Hollenbach \&
Wolfire 2006; Abel \etal 2005) of \hii\ region/PDR complexes predict
ratios of this order for Solar metallicity gas. In our observations,
the PDR origin of the $\cp$ emission is directly seen, for example, in
the map of LMC N~11, which shows that the $\cp$ emission is correlated
with the CO emission, tracing the remnant molecular gas and not the
\hii\ regions. (2) The volume of \hii\ regions is set by the
attenuation of Lyman continuum photons by absorption by neutral
hydrogen, and is independent of the metallicity.  Hence the
\hii\ region contribution in low-metallicity galaxies such as the LMC
and the SMC will be {\it less} important than it is in Galactic
star-forming regions: the $\cp$ column in the \hii\ region will
decline with metallicity $Z$, while the $\cp$ column in a PDR is
nearly independent of $Z$ (Maloney \& Wolfire 1997). As shown by
Kaufman \etal (2006), the ratio of the PDR to \hii\ region
contributions is expected to scale as $1/Z$. Hence we can safely
assume that any \hii\ region contribution is minor.

\subsection{Estimates of $G_o$}
As in our previous studies of $\cp$ emission from the Large Magellanic
Cloud, we have the advantage of much higher spatial resolution
($1'=15-17$ pc for assumed LMC and SMC distances of 50 and 60 kpc
respectively) than is generally possible for extragalactic
observations.  This is especially important since the most important
parameter controlling the physical and thermal structure of a PDR,
$G_o/n$ (where $G_o$ is the flux over the range $6-13.6$ eV,
normalised to $1.6\times 10^{-3}$ erg cm$^{-2}$ s$^{-1}$, and $n$ is
the gas density), can ordinarily only be estimated indirectly (\eg
Wolfire \etal 1989, 1990). For the LMC and the SMC, however, as we
showed in our previous study of the N~159 and N~160 regions (Paper I),
we can obtain much more direct estimates of $G_o$ using observations
of the 158\mm\ line-emitting regions at other wavelengths. 

We do this in two different ways. In the first, we obtain an estimate
of the Lyman continuum photon flux from observations of either
extinction-corrected H$\alpha$ fluxes or radio continuum observations
(Caplan \etal 1996; Ye \etal 1991; Filipovi\'c \etal 1998). There is
no indication that the IMF's of ionizing star clusters in the SMC, the
LMC and the Milky Way are grossly different; minor differences do not
influence our conclusions.  We assume a point source for the
ionising radiation and an average distance which is based on the
extent of the 158\mm\ emission. For an assumed black-body temperature
$T_{\rm BB}$, comparison of the theoretical and observed Lyman
continuum photon luminosities then gives the scaling factor to obtain
the $6-13.6$ eV flux for the observed regions. As we have seen, N~11
is excited by very luminous O stars, as are also the SMC clouds (N66:
Walborn \& Blades 1986; Massey \etal 1989; N83-N84: Testor \& Lortet
1987; for the other \hii\ regions, the ionisation parameter as derived
from the radio continuum observations also indicates that the exciting
stars are O stars), hence we assumed a temperature $T_{\rm BB}=50,000$
K. The derived values of $\go$ (for all sources for which H$\alpha$
and/or radio continuum data are available) are given in
Table~\ref{PDR}.

Since the emitting regions are spatially resolved in the infrared, we
can also use the far-infrared surface brightness to estimate $G_o$. If
we assume that all of the incident FUV radiation is absorbed by dust
within a cloud and re-radiated in the far-infrared, then the IR
surface brightness is simply related to $\go$: \be
\Iir=1.3\times10^{-4} \fuv\go \ee where $\fuv=2$ for B stars (which
emit approximately half of their flux between 6 and 13.6 eV, and the
other half at energies $E < 6 $ eV, so that the total flux absorbed by
the dust is about twice that in the $6-13.6$ eV band), and $\fuv=1$
for O stars (which also emit about half their total flux in the
$6-13.6$ eV band, but emit the other half at energies $E > 13.6$ eV;
these photons are all absorbed within the \hii\ region -- see
Hollenbach \etal 1991, hereafter HTT). The values of $\go$ derived
from the infrared surface brightness (with $\fuv=1$ since, as noted
earlier, all of these regions are powered by O stars) are also given
in Table~\ref{PDR}. The uncertainties in these estimates of $\go$ are
probably about a factor of 3.

In N~11, the two methods yield modest radiation field strengths
(except for the very bright N11AB complex) that are in good agreement
with each other. In the SMC objects we get much higher radiation field
estimates from the Lyman continuum flux than from the far-infrared
surface brightness. The IR surface brightnesses in the SMC are
anomalous in another respect as well. Simply from energy balance
considerations, it is possible to derive an equation relating the dust
temperature to $\go$ (HTT): $T_d=12.2 G_o^{0.2}$. Inverting equation
(1) to an expression for $\go$ and substitution into the expression
for $T_d$ then gives a relation between $T_d$ and the infrared surface
brightness: 
\be T_d=73\left({\Iir\over \fuv}\right)^{0.2}
\ee 
where $\Iir$ is in \ergsr; this can be rearranged to an expression for
the expected IR surface brightness as a function of dust temperature:
\be 
\log\Iir=-9.318+5\log T_d+\log \fuv\;.  
\ee 
Pak \etal (1998) derived an analogous expression (assuming $\fuv=2$)
and argued that it is obeyed by Galactic PDRs (specifically, Orion and
NGC 2024), but substantially violated in both the LMC and the SMC,
indicating that the beam filling factors in the clouds are of order
$0.1$. However, the data plotted in Pak \etal (their Figure 7) do not
really support this interpretation. If the SMC points are excluded,
the mean value of the ratio of the observed IR surface brightness to
that predicted using equation (3) (with $\fuv=1$, as appropriate for
both the LMC and SMC) is $0.7$. The mean value for the SMC points is
only $0.05$. This is completely inconsistent with beam filling factor
arguments unless clouds in the SMC are much smaller than those in the
LMC, as the expected decrease for the SMC would be only a factor of
$(50/60)^2\approx 0.7$, whereas the observed decrease is an order of
magnitude greater. In fact, the mean cloud sizes in the SMC are not
different from those in the LMC (cf.~Table \ref{PDR}).

It is quite likely, therefore, that in both the LMC and the SMC the
beam filling factor is generally close to unity (for the LMC, this
interpretation is supported by the generally good [$\sim$ factor of
two] agreement between the values of $\go$ estimated from the IR
surface brightness and those estimated from alternative methods (Paper
I and this paper), and that the discrepancy between the observed and
expected $T_d-\Iir$ relation has another cause. This is almost
certainly the reduced dust-to-gas ratio in the Magellanic Clouds,
especially the SMC. The ratio of (total) hydrogen column $N_H$ to
visual extinction $A_V$ for the (Solar neighborhood) Milky Way, the
LMC, and the SMC are $N_H/A_V=1.9\times 10^{21}$, $8.0\times 10^{21}$,
and $(1.7-6.7)\times 10^{22}\;\psqcm$ mag$^{-1}$, where the Galactic
value is from Bohlin \etal (1978), the LMC value is from Fitzpatrick
(1985), and the lower and upper ends of the SMC range are from Bouchet
\etal (1985) and Schwering (1988), respectively. A cloud of nominal
column $N_H=2\times 10^{22}\;\psqcm$ in the SMC therefore has a visual
extinction $A_v\sim 1$, so that the PDR (whose extent is determined by
the absorption of the FUV photons by dust) may occupy the entire
column, and some fraction of the incident FUV photons will pass
through the entire column without absorption and escape the galaxy
entirely. The dramatic overall decrease in IR surface brightness in
the SMC (for a fixed dust temperature) even compared to most of the
LMC points suggests that this is in fact the case; as we show below,
quantitative analysis of the $\cp$ 158\mm\ emission from the SMC lends
further support to this argument.

\subsection{PDR parameters}
\nobreak The \cii\ 158\mm\ line intensity is directly proportional to
the $\cp$ column density if the emission is optically thin. The {\it
  minimum} value for this column follows from the assumption that the
emission occurs in the high-density ($n\gg n_{\rm cr}\approx 3000
\pcubcm$), high-temperature ($T \gg 91$ K) limit. For a resolved
source we find:
$$\Icii = 5\times10^{-4} N(\cp)[17.5]\; \ergsr\eqno(4)$$ and 
$$\taucii = 0.163 N(\cp)[17.5] \Delta V_{5}^{-1} (91/T)\eqno(5)$$
where $N(\cp)[17.5]$ is the $\cp$ column density in units of
$3\times10^{17}\psqcm$, and $\Delta V_{5}$ is the line FWHM in units
of 5 $\kms$. In LMC \cii\ clouds, $\Delta V_{5}\,=\,2$ (Boreiko \&
Betz 1991). From this we obtain the minimum hydrogen PDR column
density $\nhmin$ (PDR):
$$\nhmin {\rm (PDR)} = 2\times10^{20}\, {\Icii\over 10^{-4}}
\,{3\times 10^{-4}\over \xcp}\psqcm\eqno(6)$$ where $\Icii$ is in
units of $\ergsr$, and $\xcp$ is the gas-phase C$^+$ abundance
relative to hydrogen. We assume that in a PDR the C$^+$ abundance
equals the total gas-phase carbon abundance, and we assume carbon
locked up in dust to be a negligible amount. The gas-phase carbon
abundances are $\xc = 8\times 10^{-5}$ and $\xc = 3\times 10^{-5}$ for
the LMC and SMC respectively, factors of about four and ten below the
Solar Neighbourhood carbon abundance (see Pagel 2003).  Thus, for LMC
conditions, equation (3) becomes:
$$\nhmin ({\rm LMC})=8\times10^{20}\;{\Icii\over  10^{-4}}\psqcm\,.\eqno(7a)$$ 
whereas for SMC conditions we should take
$$\nhmin ({\rm SMC})=2\times10^{21}\;{\Icii\over 10^{-4}}\psqcm\,.\eqno(7b)$$

Both the N~11 results presented here as well as the N~159/N~160 (Paper
I) and 30 Dor (Poglitsch et al. 1995) results show that in the LMC
\cii\ clouds, the detected 158\mm\ emission ranges from about
$7\times10^{-5}\;\ergsr$ up to about $4\times10^{-4}\; \ergsr$, so
that the PDR column densities range from $N_H = 6\times 10^{20}\psqcm$
up to (beam-averaged) peaks of $N_H = 3\times 10^{21}\psqcm$. Although
the 158\mm\ emission detected from the SMC is about {\it three times
weaker}, the {\it lower} carbon abundance compensates for this, and
beam-averaged PDR column densities are $N_H =
0.5-4\,\times\,10^{21}\psqcm$, essentially identical to the LMC
results.

Under realistic conditions of finite density and temperature, the true
column densities are typically double these values (Stacey \etal
1991), a conclusion which is supported by our estimate in section
3.4.3.  The minimum PDR masses are calculated from the integrated
158\mm\ emission, assuming that the emission is optically thin, an
assumption which we have shown to be correct in Paper I. The procedure
is as follows. We obtain the effective emitting area $\aeff$ by
dividing the integrated 158\mm\ emission by the observed peak
intensity. This peak intensity is inserted in equations (7) to yield
$\nhmin$. We then convert $\nhmin\times\aeff$ into a gas mass
$\mhmin$, including a factor 1.3 to correct for helium.  The results
are given in Table \ref{PDR}. Again, the actual PDR masses may be a
factor of two or more higher than the calculated $\mhmin$ in the more
realistic case of finite temperature and density.

In Table 6, we have also included the `virial mass', calculated from
the Galactic relation $\mvt = 39 L_{\rm CO}^{0.8}$ (Solomon \etal
1987; but see also Maloney 1990).  We find that the PDR masses
suggested by the \cii\ luminosities are very similar to these $\mvt$
masses (derived under the assumption of Milky Way conditions) in the
case of the LMC, whereas they significantly exceed these CO-based
masses in the SMC.  In addition, this effect is most pronounced, {\it
  in either Magellanic Cloud}, for the very brightest \hii\ regions:
30 Doradus in the LMC ($\mpdr/\mvt$ = 2.8) and N~66 in the SMC
($\mpdr/\mvt$ = 5.7).  This is in striking contrast to the
extragalactic sample of Stacey \etal (1991), where the derived ratios
of $\mhmin/M_{\rm H_2}$ (obtained using a `standard' conversion factor
for CO, and are equivalent to our $\mpdr/\mvt$) are typically a few
\%, with a maximum of about 0.1. The LMC and SMC clouds observed with
the KAO all exceed these values by one to two orders of
magnitude. This remarkable difference suggests that in the LMC and the
SMC molecular clouds are considerably under-luminous in CO for their
masses, and that in addition in the brightest \hii\ regions (30
Doradus, N66) a rather large fraction of the original molecular cloud
mass has already been converted into a PDR (cf.  Poglitsch \etal
1995).  Indeed, in the LMC CO appears to be under-luminous by a factor
of 3-6, and in the SMC by a factor of 15-60 (Israel \etal 1993; Israel
1997) so that actual molecular gas masses are expected to exceed the
CO-derived `virial masses' by a similar factor. Thus, the KAO
\cii\ measurements indicate the presence of significant amounts of
hydrogen, most likely in the form of $\h2$, not traced by CO.  A
similar situation has been noted to occur in the magellanic irregular
dwarf galaxy IC~10, where \cii\ observations likewise points at such
an $\h2$ reservoir (Madden \etal 1997). The \cii\ results confirm
those derived by using thermal continuum emission from dust grains as
a tracer for total (atomic + molecular) gas masses in the Magellanic
Clouds (Israel, 1997; Leroy \etal 2007, 2009).

In Paper I, we have shown that for all temperatures $T > 11$ K, the
\cii\ emission from the observed LMC sources is optically thin
($\taucii\,<<\,1$).  By the same reasoning, this is also true for N~11
as well as the SMC clouds discussed in this paper.  We also derived
{\it a lower limit} to the \cii\ excitation temperature by assuming
the \cii\ emission to be {\it optically thick} instead:
$$\texc=91.2\left[\ln\left({19.2\Delta V_5\over
    I_{-4}}+1\right)\right]^{-1}\; {\rm K}\eqno(8)$$ where
$10^{-4}I_{-4}\;\ergsr$ is the integrated intensity of the
158\mm\ line. The peak intensities in N~11 and the SMC clouds are
$I_{-4} =0.3-2.2$, and so, with $\Delta_5 \approx\,2$ (Boreiko \& Betz
(1991)), eqn.~(8) gives $\texc\,=\,20-25$ K.  Again, these are lower
limits to the actual gas kinetic temperature for two reasons: (a) the
line is actually optically thin, and (b) the line may be sub-thermally
excited. The densities in the emitting regions are probably high
enough that the $\cp$ fine-structure levels are close to thermal
equilibrium.

\subsection{Photoelectric heating efficiencies}

The results presented in this paper confirm the conclusion from Paper
I that the efficiency of grain photoelectric heating, as inferred from
the measured values of $\fcii/\fir$ (Tables \ref{lmcdat} and
\ref{smcdat}) is remarkably high for the observed LMC objects, with
typical values of $1-5\%$, and somewhat lower but still high in the
SMC, with typical values of $0.5-2\%$. We may use the inferred heating
efficiency together with our earlier estimates of $\go$ to estimate the
gas densities in the emitting regions. In Paper I we have shown that
the idealised analytical fit for the photoelectric heating efficiency
$\epsilon$ as a function of $\go$, electron density $n_e$, and gas
temperature $T$ given by Bakes \& Tielens (1994),
$$\epsilon={3\times 10^{-2}\over 1+2\times 10^{-4}\go
  T^{1/2}/n_e}\;.\eqno(9)$$ may be rewritten as an expression for the
total hydrogen density, $n_o$:
$$n_o=0.002\go x_C^{-1}
T_2^{1/2}(3/\epsilon_{-2}-1)^{-1}\pcubcm\eqno(10)$$ 
where $\epsilon_{-2}=\epsilon/10^{-2}$, $x_C$ is the carbon abundance,
and $T_2=T/100$, assuming that all electrons are supplied by carbon,
and all carbon in the PDR zone is ionised. After substitution of the
actual values of $\go$, $x_C$, and $\epsilon_{-2}$, this simple model
yields large values $n_o/T_2^{1/2}\,=\,10,000-20,000\,\pcubcm$
K$^{-1/2}$ for N159E, N159W. Significantly lower values
$n_o/T_2^{1/2}\,=\,2100\pm400\,\pcubcm$ K$^{-1/2}$ are found for N160
and the N11 clouds. The $\go$ values derived for the SMC have a large
spread which is reflected by $n_o/T_2^{1/2}$ ranging from
$800\,\pcubcm$ K$^{-1/2}$ to $\approx\,20,000\,\pcubcm$ K$^{-1/2}$
with a mean of $\sim\,4000\,\pcubcm$ K$^{-1/2}$. As $T_2$ is probably
within a factor of two of unity, the implied range of gas densities is
reasonable.  In the case of N~159 we can verify this because Pineda
\etal (2008) have also observed the region in various (sub)millimeter
emission lines. Their detailed PDR modelling has yielded temperatures
of $\sim 80$ K and densities of $10^{4}\,\pcubcm$, very close to the
more crude estimate we have presented above.

Thus, there is no need to invoke the actually-existing differences in
grain properties between the LMC, the SMC, and the Galaxy in order to
explain the high photoelectric heating efficiencies. We note that some
of our LMC $\fcii/\fir$ values exceed the maximum $\epsilon\,=\,0.03$
implied by eqn.~8. This is not a problem, as Bakes \& Tielens (1994)
show that more realistic models including small grains and polycyclic
aromatic hydrocarbons (PAHs) allow higher heating efficiencies $\geq
0.05$ for low values of $\go$ (cf.~their Fig.~13). The primary reason
for the exceptionally large values of $\epsilon$ observed in the SMC
and especially in the LMC is the relatively low values of $\go$
characterising the emitting regions.

\section{Discussion and conclusions}

The Large and Small Magellanic Clouds present unique opportunities for
studying star formation and the interstellar medium in environments
very different from our own, with reasonably high spatial resolution.
The observations presented here and in Paper I of a sample of bright
\hii\ regions in the LMC and the SMC highlight marked differences in
the properties of the interstellar medium associated with these
star-forming complexes. These differences result primarily from the
lower metal abundances and dust-to-gas ratios of the SMC and the LMC
with respect to the Galaxy and to each other.

All of the star-forming regions observed in the Magellanic Clouds show
158\mm\ line emission significantly enhanced relative to CO $J=1-0$
emission ($I_{CII}/I_{CO}\,=\,1.5-10\,\times10^4$) as compared to the
emission arising from star-forming molecular clouds in our own Galaxy
($I_{CII}/I_{CO}\,=\,0.1-1.4\,\times\,10^4$), or in the nuclei of
IR-bright galaxies ($I_{CII}/I_{CO}\,=\,0.1-0.8\,\times\,10^4$ - see
Stacey \etal 1991; Fixsen \etal 1999). There appears to be no
significant difference between the LMC and the SMC in this respect.
This behavior is caused by the relative constancy of the total $\cp$
column in a PDR if both the gas-phase carbon abundance and the
dust-to-gas ratio are varied in the same way (\eg van Dishoeck \&
Black 1988; Boreiko \& Betz 1991; Maloney \& Wolfire 1997), in
contrast to a decrease in the CO column density when self-shielding is
important (Maloney \& Black 1988; van Dishoeck \& Black 1988; Maloney
\& Wolfire 1997). In effect, in a low-metallicity molecular cloud the
size of the CO-emitting core will shrink, so that the PDR will occupy
a larger fraction of the total cloud volume (cf.~Israel \etal 1986;
Israel 1988; Bolatto et al. 1999; R\"ollig \etal 2006; Shetty \etal
2010). This is illustrated by the fact that the minimum masses in the
PDR (calculated in the high-density, high-temperature limit) are a
much larger fraction, by an order of magnitude, of the associated
molecular gas mass (as inferred from CO) for the Magellanic Cloud
regions than in Galactic star-forming regions. Thus, either these
clouds are very under-luminous in CO compared to Galactic molecular
clouds, or a much larger fraction of the mass of the molecular cloud
has been photo-dissociated. The observation that the most intense
\hii\ regions in either Cloud (30 Doradus, N66) exhibit this effect
most strongly suggests that both effects occur in tandem.

We have also established that the ratios of the 158\mm\ line to the
far-infrared continuum -- a measure of the efficiency of grain
photoelectric heating -- in the LMC and the SMC are unique in the
sense that they are consistently higher than those seen
elsewhere.  They exceed, again by an order of magnitude, the
$\fcii/\fir$ ratios seen in the star-forming regions of the Milky Way
and M~33 (Higdon \etal 2003), although we should note that some
star-forming regions in M~51 (Nikola \etal 2001) and M~31
(Rodriguez-Fernandez \etal 2006) have ratios
($\fcii/\fir\,=\,1-2\,\%$) similar to those in the SMC.  The global
$\fcii/\fir$ ratios of the Milky Way (Fixsen \etal 1999) and indeed
those of some 60 other galaxies observed by ISO (Malhotra \etal 2004;
Negishi \etal 2004) are much lower, with $\fcii/\fir\,=\,0.2-0.8\,\%$.
The large implied efficiencies, of the order of 1--2\%, can be
understood as the result of relatively normal PDR gas densities
($n_o\sim 10^3-10^4$) combined with unusually low values of the
ambient FUV photon flux, $\go\,=\,30-350$. The underlying cause of
such low values of $\go$ again is low metallicity and dust content of
the Clouds, which provides UV photons with a relatively long mean free
path length.  In such environments, the sphere of influence of an FUV
photon source is much larger than in environments with solar
metallicities, and the geometric dilution of the radiation field is
correspondingly larger.

Although one might expect that the more metal-poor SMC would have even
higher $\fcii/\fir$ ratios than the LMC, this is not the case; they
are on average a factor of 2-3 lower.  We also note that the $\go$
values derived for the SMC from emission by ionised gas are a factor
of 3-4 higher than those derived from emission by dust whereas the two
are essentially the same for the LMC.  As the \cii\ heating appears to
be dominated by PAH particles (Helou \etal 2001; Rubin \etal 2009), we
ascribe these differences to the combined effect of greater geometric
dilution of the radiation field which increases {\it gas heating
  efficiencies}, and at the same time lower {\it gas heating rates} by
the very low PAH abundances in the SMC (Sandstrom \etal 2010). The
galaxy IC~10 has metallicities and PAH abundances in between those of
the LMC and SMC (see Wiebe \etal 2011). Its \cii-related ratios
(Madden \etal 1997) are indistinguishable from those of the LMC and
SMC (Fig.\,\ref{ratio}.  This suggests that the combined effects of
dust and PAH abundances cause low-metallicity galaxies to have rather
similar \cii\ , CO and FIR ratios largely independent of actual
metallicity.

\acknowledgements It is a pleasure to thank the MPE FIFI personnel,
notably Sue Madden, Albrecht Poglitsch and Gordon Stacey, without
whose generous help and support the observations described in this
paper could not have been obtained.  We also thank the KAO crew for
their unfailing support under sometimes difficult flight conditions.
P.R.M. acknowledges support through the NASA Long Term Astrophysics
Program under grant NAGW-4454, support from the NSF through grant
AST-0705157 and support from NASA through grant 1394366. He would also
like to thank the Netherlands Organisation for Pure Research (N.W.O.)
for support for visits to Leiden to work on this project.

\end{document}